\documentclass[a4paper,reqno]{article}

    \usepackage[english]{babel} 

    \usepackage{graphicx}
    \usepackage[]{color}  
    \usepackage[dvipsnames]{xcolor} 
    \usepackage{csquotes} 
    \usepackage{appendix} 
    \usepackage[a4paper]{geometry} 
    \usepackage{cancel} 
    \usepackage{tikz}
    \usepackage{enumitem}
    \usepackage{stmaryrd}
        \usetikzlibrary{backgrounds}

    \usepackage{amsmath,amssymb,amsfonts,amsthm} 
    \usepackage{mathtools,todonotes} 
    \usepackage{tikz-cd}  
    \usepackage[all,cmtip]{xy} 
    \usepackage[bbgreekl]{mathbbol} 
    \usepackage{array} 

    \usepackage{authblk} 
    \usepackage{hyperref} 

\makeatletter
\newcommand{\subjclass}[2][1991]{%
  \let\@oldtitle\@title%
  \gdef\@title{\@oldtitle\footnotetext{#1 \emph{Mathematics subject classification.} #2}}%
}
\newcommand{\keywords}[1]{%
  \let\@@oldtitle\@title%
  \gdef\@title{\@@oldtitle\footnotetext{\emph{Key words and phrases.} #1.}}%
}
\makeatother

\theoremstyle{definition} 
    \newtheorem{definition}{Definition}

\theoremstyle{plain} 
    \newtheorem{theorem}[definition]{Theorem}
    
    \newtheorem{lemma}[definition]{Lemma}

\theoremstyle{remark} 
    \newtheorem{remark}[definition]{Remark}

\usepackage[bibstyle=alphabetic,citestyle=alphabetic,useprefix,giveninits=true, sorting=ynt, sortcites, minbibnames=99,maxbibnames=99,backend=biber]{biblatex}  

\renewbibmacro{in:}{} 
\DeclareSourcemap{
  \maps[datatype=bibtex]{
    \map[overwrite]{ 
      \step[fieldsource=doi, final]
      \step[fieldset=url, null]
      \step[fieldset=eprint, null]
      }
     \map[overwrite]{ 
      \step[fieldsource=eprint, final]
      \step[fieldset=pages, null]
      \step[fieldset=eid, null]
      \step[fieldset=journal, null]
    }  
  }
}

    \newcommand{\ugam}{\underline{\gamma}}
    \newcommand{\igam}{\iota_{\hat{\gamma}}}
    \newcommand{\vf}{\iota_\varphi}

    \newcommand{\qsp}[2]{\,\ensuremath{\raise.5ex\hbox{$#1$}\big\slash\raise-.5ex\hbox{$#2$}}}

    \makeatletter
        \newcommand{\zzlabel}[1]{\ifmeasuring@\else\ltx@label{#1}\fi} 
    \makeatother

    \newcounter{terms}[equation] 
    \newcommand{\unl}[2]{\underline{#1}_{\refstepcounter{terms} \zzlabel{#2} \theterms}} 

    \newcommand{\reft}[2]{(\ref{#1}.\ref{#2})} 

\title{BV description of $N=1, D=4$ Supergravity in the first order formalism }
\author[1]{Alberto S. Cattaneo}
\author[1,2]{Filippo Fila-Robattino}

\renewcommand\footnotemark{}

\affil[1]{Institut f\"ur Mathematik, Universit\"at Z\"urich }
\affil[2]{Scuola Internazionale Superiore di Studi Avanzati, Trieste} 
\thanks{ASC acknowledges partial support of the SNF Grant No. 200021 227719 and of
the Simons Collaboration on Global Categorical Symmetries. This research was
(partly) supported by the NCCR SwissMAP, funded by the Swiss National Science Foundation.
This article is based upon work from COST Action 21109 CaLISTA, supported by COST (European Cooperation in
Science and Technology) (www.cost.eu), MSCA-2021-SE-01-101086123 CaLIGOLA, and MSCA-DN CaLiForNIA
-101119552. FFR acknowledges funding from the EU project Caligola HORIZON-MSCA-2021-SE-01, Project ID: 101086123.}
\date{}                     

\begin{document}

\maketitle
\begin{abstract}
   This note examines the BV formulation of $N=1$, $D=4$ supergravity in the first-order Palatini--Cartan framework. Challenges in achieving an off-shell formulation are addressed by introducing corrections to the rank--2 BV action, offering in addition a solid foundation for the study of the theory on manifolds with boundary. 
\end{abstract}

\tableofcontents

\section{Introduction}
Supergravity is the supersymmetric extension of general relativity, where local supersymmetry transformations are included as part of a larger symmetry algebra. In particular, $N = 1$, $D = 4 $ supergravity represents the case in which a single supersymmetry generator is introduced in four dimensions. This paper deals with achieving the off-shell supersymmetry within the BV formalism, generalizing the work of \cite{Baulieu:1990uv} in a way that allows to account for possible boundary extensions in the context of the BV/BFV formalism.

The B(F)V algorithm was first introduced in \cite{BV1,BV2,BV3} as a way to deal with the quantization of systems with gauge symmetries, extending the construction provided by the BRST formalism to more general cases. Cattaneo, Mnev and Reshetikin \cite{ CMR2012b, CMR2012, CMR2} later refined the work of Batalin, Fradkin and Vilkovisky, considering the case of manifolds with boundary, where one is required to work with theories in the first order formalism.

In the case of supergravity, this equates to studying the theory in the Palatini--Cartan formalism, allowing the spin connection to be a dynamical field. Therefore, a supersymmetry transformation for the spin connection needs to be derived by imposing the invariance (up to boundary terms) of the classical action, obtaining a non-vanishing expression. 
As expected, when squaring the supersymmetry one obtains, up to equations of motion, the action of the diffeomorphism symmetry, whose gauge parameter depends quadratically on the generator of the SUSY. 

Collecting the action of all the symmetries in a single operator $Q_0$, the above statement is equivalent to $Q_0^2\approx0$, where the symbol $\approx$ indicates an equality only on--shell. The BV procedure requires the introduction of ghosts, seen as degree 1 local generators of the symmetries, and anti--fields, which differentiate into field momenta and ghost momenta respectively of degree -1 and -2, obtaining a $\mathbb{Z}$--graded symplectic supermanifold which takes the role of space of BV fields. To such space, the BV algorithm then assigns an action functional $\mathcal{S}$ of degree 0, whose Hamiltonian vector field $Q$ extends $Q_0$ and is cohomological, achieving, in the present case, the off--shell closure of supersymmetry. The nilpotency of $Q$ is equivalent to the so--called 'Classical Master Equation', which is the requirement that $\mathcal{S}$ Poisson commutes with itself, with respect to the canonical Poisson structure induced by the BV symplectic form. In order to obtain such property, typically one extends the classical action by terms which contain the anti--fields, encoding the symmetries of the system. 

Contrary to the case of pure Palatini--Cartan gravity \cite{Piguet:2000fy, Moritsch:1993eg, Baulieu:1985md}, where the BV action is shown to be linear in the anti--fields, in the case of $N=1,D=4$ supergravity it was found \cite{Baulieu:1990uv} that the BV action is of rank 2, i.e. quadratic in the anti--fields. It is also the case in the present work, where, starting from the pure gravity case studied in \cite{CS2017}, we find the BV action to be 
    \begin{align*}
        \mathcal{S}_{SG}&=\int_M \frac{e^2}{2}F_\omega + \frac{1}{3!} e \bar{\psi}\gamma^3 d_\omega\psi -(L_\xi^\omega e - [c,e] + \bar{\chi}\gamma\psi) e^\dag -i (L_\xi^\omega \bar{\psi} - [c,\bar{\psi}] -d_\omega \bar{\chi})\psi_\dag\\
        &\qquad + (\iota_\xi F_\omega - d_\omega c + \delta_\chi \omega)\omega^\dag   + \left(\frac{1}{2}\iota_\xi\iota_\xi F_\omega -\frac{1}{2} [c,c] + \iota_\xi \delta_\chi\omega  \right)c^\dag+\frac{1}{2}(\iota_{[\xi,\xi]} + \iota_\varphi)\xi^\dag\\
        &\qquad   - i\left(L_\xi^\omega\bar{\chi} - [c,\bar{\chi}]  -\frac{1}{2}\iota_\varphi \bar{\psi} \right)\chi_\dag +\frac{1}{2}\left(\check{\omega}-\frac{1}{2}e\iota_\xi \check{c} - \check{c}\iota_\xi e\right)\vf e^\dag
    \end{align*}
    \begin{align*}
        &\qquad +\frac{1}{4}\left( \frac{1}{2}\bar{\psi}^0_\dag\ugam + \alpha(\check{\omega}\bar{\psi})\ugam -\frac{i}{2}\iota_\xi\check{c}\bar{\psi}-\alpha(\check{c}\iota_\xi e \bar{\psi})\ugam -\frac{i}{2}\check{c}\bar{\chi}\right)\vf\psi_\dag\\
        \nonumber&\qquad + \frac{i}{4\cdot 3!}\left(\frac{1}{2}\alpha(\check{\omega}\bar{\psi})\ugam -\frac{i}{2}\iota_\xi\check{c}\bar{\psi}-\alpha(\check{c}\iota_\xi e \bar{\psi})\ugam -\frac{i}{2}\check{c}\bar{\chi}\right)\gamma^3\vf(\check{\omega}\psi)\\
        &\qquad -\frac{i}{2\cdot 3!}\left( \frac{1}{2}\bar{\psi}^0_\dag\ugam+\frac{1}{2}\alpha(\check{\omega}\bar{\psi})\ugam -\frac{i}{2}\iota_\xi\check{c}\bar{\psi}-\alpha(\check{c}\iota_\xi e \bar{\psi})\ugam \right)\gamma^3\chi <e,\bar{\chi}[\check{\omega},\gamma]\psi>\\
        \nonumber&\qquad +\frac{1}{2\cdot 3!}\left( \frac{1}{4}\bar{\psi}^0_\dag\ugam-\frac{i}{2}\iota_\xi\check{c}\bar{\psi}-\alpha(\check{c}\iota_\xi e \bar{\psi})\ugam \right)\gamma^3\chi <e,\bar{\chi}\ugam^2\psi^0_\dag>\\
        \nonumber&\qquad -\frac{1}{32}\left(i\bar{\psi}_\dag \chi +\frac{1}{3!}(\check{\omega} - e \iota_\xi \check{c} - 2 \check{c}\iota_\xi e  )\bar{\psi}\gamma^3 \chi \right)\bar{\chi}\igam\igam([\check{\omega},\gamma]\psi)\\
        \nonumber&\qquad - \frac{i}{32}\left(i\bar{\psi}_\dag\chi + \frac{1}{3!}(e\iota_\xi \check{c} + 2\check{c}\iota_\xi e )\bar{\psi}\gamma^3\chi  \right)\bar{\chi}\igam\igam(\ugam^2\psi^0_\dag),
    \end{align*}
where the definitions of the fields, anti--fields and implicit expressions are given in chapter \ref{sec: BV Sugra}.

The study of supergravity in the first--order formalism, while producing cumbersome expressions, is the correct starting point for the BV/BFV analysis in the case of manifolds with boundary, which will allow to obtain the reduced phase space of the theory \cite{CMR2012b, CMR2012, CMR2} in a way compatible with quantization.  In particular, a work in this direction is in progress \cite{CFR25b}, following the lines of \cite{CCS2020}.

In a recent paper \cite{grigoriev2025presymplecticbvakszn1d4}, the BV action for the half-shell formalism has been recovered from a presymplectic version of the AKSZ formalism. It would be interesting to see if their formalism could be adapted to recover our Palatini--Cartan version.
    
\section{Review of the formalism and setting}
\subsection{Palatini-Cartan SUGRA}
Supergravity is defined as the supersymmetry theory containing gravity, in which the SUSY is realised locally (the spinor parameter $\chi$ is a function of the spacetime coordinates $\chi(x)$. We investigate here the $\mathcal{N}=1$ case, namely the case in which only one supersymmetry generator is introduced, in 4 dimensions, as it is the starting point for further generalizations.
We start with pure gravity, and subsequently couple it with a Majorana-type spinor, which will act as the gravitino, the superpartner of the graviton. 

Let $M$ be a spin manifold and let $P_{\text{Spin}}$ be a principal Spin$(3,1)$ bundle over $M$.  We introduce a 4-dimensional real vector space $V$ with a Lorentz-type metric $\eta$ of signature $(-,+,+,+)$. Without loss of generality we can assume that $\eta=$diag$(-1,1,1,1)$ is the Minkowski metric and define the associated bundle (called 'Minkowski bundle') $\mathcal{V}:=P_{\text{Spin}} \times_\Lambda V$, where $\Lambda$ is the spin 1 representation of Spin$(3,1)$. 
\begin{remark}\label{rem: SO spin split}
    Notice that the double cover $l\colon$Spin$(3,1)\rightarrow SO(3,1)$ induces a bundle morphism to a $SO(3,1)$ bundle $\hat{l}\colon P_{\text{Spin}} \rightarrow  P_{SO} $, hence $\mathcal{V}\simeq P_{SO} \times_{\Lambda_0} V$, where $\Lambda_0$ is the vector representation of $SO(3,1)$, such that $\Lambda=\Lambda_0\circ l$. Furthermore, one can identify elements of the Lie algebra of Spin$(3,1)$ with the second wedge power of $V$, as it defines $4\times 4$ antisymmetric  matrices: $\mathfrak{spin}(3,1)=\mathfrak{so}(3,1)\simeq \wedge^2 V $.
\end{remark}
The last ingredient we need in our setting is what is commonly known as Dirac spinor bundle, namely the following associated vector bundle $\mathbb{S}_D:=P_{\text{spin}}\times_\gamma \mathbb{C}^4 $, where $\gamma$ is the gamma representation of the Clifford algebra $\mathcal{C}(V)$ restricted to its spin subgroup Spin$(V)\simeq$Spin$(3,1)$.\footnote{For more details about the notations and the convention see \cite{FR25}}

The independent fields of the theory are:
\begin{itemize}
    \item The coframe $e$ (also known as vielbein or tetrad in $D=4$) defined as an isomorphism $e:TM\rightarrow \mathcal{V}$, inducing a metric on spacetime as $g:=e^*(\eta)$, i.e. such that $g_{\mu\nu}=e_\mu^a e_\nu^b \eta_{ab}$, where $\mu=1,2,3,4$ are curved indices on $M$ while $a=0,1,2,3$ are flat indices on $V$.\footnote{Note that $e$ enjoys an internal Lorentz symmetry (acting on the flat indices) on top of the usual diffeomorphisms.} The coframe has the advantage of being expressed as a differential form, indeed $e=e_\mu^a dx^\mu v_a\in\Omega^1(M,\mathcal{V})$, where $x$ are coordinates on $M$ and $\{v_a\}$ is a basis of $V$.
    \item  The spin connection $\omega$. The space of connections is denoted by $\mathcal{A}_M$, and is locally modeled by 1-forms on $M$ with values in the Lie algebra $\mathfrak{so(3,1)}=\mathfrak{spin(3,1)} $, in our notation $\omega=\omega_\mu dx^\mu v_a\wedge v_b \in\Omega^1(M,\wedge^2\mathcal{V})$.
    \item The gravitino $\psi$, a spin-$\frac{3}{2}$ Majorana spinor, i.e. a 1-form on $M$ with values in the subbundle of Majorana spinors $\mathbb{S}_M\coloneqq\{ \chi \in \mathbb{S}_D \hspace{1mm}\vert\hspace{1mm} \bar{\chi}:=\chi^\dagger \gamma_0 = \chi^t C \} $, where $C$ is the charge conjugation matrix. Furthermore, as we are dealing with a fermion, we need to reverse the parity\footnote{The parity reversed Majorana spinor bundle is defined as $\Pi \mathbb{S}_M$ and simply given by $\mathbb{S}_M$ with the requirement that the components of each spinor are Grassmann-odd.} of $\mathbb{S}_M$, obtaining $\psi=\psi_\mu dx^\mu \in \Omega^1(M,\Pi \mathbb{S}_M)$.
\end{itemize}

The theory is described by the following action functional\footnote{We omit the symbol $\wedge$ when multiplying differential forms and sections of the exterior algebra of $V$, but the wedge product is assumed in both. Parity in the algebra is defined as the sum of the fermionic parity, the form degree modulo $2$,
the degree in $\Lambda V$ modulo $2$, and the ghost number (to be introduced below) modulo $2$.}
\begin{equation}
    S_{SG}=\int_M \frac{e^2}{2}F_\omega + \frac{1}{3!}e \bar{\psi}\gamma^3 d_\omega \psi,
\end{equation}
where $F_\omega=d\omega + \frac{1}{2}[\omega,\omega]$ is the curvature of the connection, $\gamma$ is an element of $V\otimes \mathcal{C}(V)$ defined by $\gamma=\gamma^a v_a$,\footnote{Notice in our notation we have the following relations
\begin{equation*}
    \{\gamma_a,\gamma_b\}=-2 \eta_{ab} \qquad \{\gamma_\mu,\gamma_\nu\}=-2 g_{\mu\nu},
\end{equation*}
having set $\gamma_\mu=e_\mu^a \gamma_a$.} and $\{v_a\}$ is a basis of $V$. Lastly, we define $d_\omega \psi \coloneqq d\psi -\frac{1}{4} \omega^{ab}\gamma_{ab}\psi $,\footnote{Alternatively, one can define for all $\alpha\in\wedge^2 V$, $[\alpha,\psi]\coloneqq\frac{1}{4}\gamma^{ab}\iota_{v_a}\iota_{v_b}\alpha\psi=-\frac{1}{4}\gamma^{ab}\alpha_{ab}\psi$, having set $\iota_{v_a}v_c\coloneqq\eta_{ac}$.} having set $\gamma_{ab}=\gamma_{[a}\gamma_{b]}=\frac{1}{2}[\gamma_a,\gamma_b] $.

\begin{remark}
    The bracket $[\cdot,\cdot]$ is defined to encode any (possibly graded\footnote{In our convention, the parity of an element $\alpha\in \Omega^i(M,\wedge^j V)$ is defined to be $|\alpha|=i+j $ mod 2. In the same way, a pure Majorana spinor has parity 1, so that in the case of the gravitino, $|\psi|=1+1$ mod $2=0$.}) Lie algebra action.\footnote{The bracket $[\cdot,\cdot]$ on $\wedge^\bullet V$ (encoding the action of the Lorentz group) can also be induced from the pairing in $V$, indeed if for any $A,B \in V$ we define $[A,B]:=-(-1)^{|B|}\eta(A,B)= -(-1)^{|B|}A^aB^b \eta_{ab}$, then one can extend the action bi-linearly to $\wedge^k V$ requiring that the graded Leibniz rule holds. Furthermore, notice that the bracket defined above is graded, i.e. $[A,B]=-(-1)^{|A||B|}[B,A] $, where $|\cdot|$ denotes the parity.}
    In the general case, if a field $\phi$ transforms in a representation $\rho$ of the Spin group, then we have $[\omega,\phi]\coloneqq\rho(\omega)(\phi)$. In the case of the gravitino field, transforming in the gamma representation, we obtain
        \begin{equation*}
            [\omega,\psi]=\gamma(\omega^{ab}v_a\wedge v_b) (\psi)= \omega^{ab}\gamma(v_a\wedge v_b)(\psi) = -\frac{1}{4} \omega^{ab}\gamma_{ab}\psi,
        \end{equation*}
    where $\gamma(v_a\wedge v_b)=-\frac{1}{4}\gamma_{ab}$ is the image under the gamma representation of the generators of the Lie algebra $\mathfrak{spin(3,1)}$\footnote{One can show $-\frac{1}{4}\gamma_{ab}$ are generators of $\mathfrak{spin(3,1)}$}.
    
    One must also be particularly careful when computing $[\omega,\gamma]$, as $\gamma$ has values in $V\otimes \mathcal{C}(V)$, namely it transforms as a Lorentz vector and via the action of the gamma representation on gamma matrices: indeed one obtains the following splitting 
        \begin{equation}
            [\omega,\gamma]=[\omega,\gamma]_V + [\omega,\gamma]_S,
        \end{equation}
    where $[\omega,\gamma]_V := \omega^{ab}\eta_{bc} \gamma^c v_a$ and $[\omega,\gamma]_S:=\omega^{ab}[\gamma(v_a\wedge v_b),\gamma^c] v_c = -\frac{1}{4}\omega^{ab}(\gamma_{ab}\gamma^c - \gamma^c\gamma_{ab} ) v_c$. It is a quick computation to show that $[\omega,\gamma]=0$ for all $\omega\in \wedge^2 V$, and therefore $d_\omega \gamma=0$ as the gamma matrices are constant.
\end{remark}

    The variation of the $\mathcal{N}=1, D=4$ supergravity action produces a boundary term and a bulk term containing the Euler-Lagrange equations
    \begin{align*}
        \delta S_{SG}&=\int_M \left(e  F_\omega + \frac{1}{3!}\bar{\psi}\gamma^3 d_\omega \psi \right)\delta e + e\left(d_\omega e -\frac{1}{2}\bar{\psi}\gamma\psi\right)\delta \omega + \frac{1}{3}\left(\frac{1}{2} d_\omega e \bar{\psi}\gamma^3 + e d_\omega \bar{\psi}\gamma^3 \right)\delta\psi\\
        &\quad - \int_{\partial M} \frac{e^2}{2}\delta \omega + \frac{1}{3!}e\bar{\psi}\gamma^3 \delta \psi,
    \end{align*}
having used the fact that\footnote{This identity is quickly obtained by applying formula \eqref{id: action of omega}. }   
    \begin{equation}\label{id: i bar gamma psi}
        -\frac{1}{3!}e\bar{\psi}\gamma^3[\delta \omega,\psi]= - \frac{1}{2}e \bar{\psi}\gamma\psi\delta \omega.
    \end{equation}
We then obtain the following equations of motion:
    \begin{eqnarray}
        & e F_\omega + \frac{1}{3!}\bar{\psi}\gamma^3 d_\omega\psi =0,\\
        & \label{eq:torsion bulk} e \left( d_\omega e - \frac{1}{2}\bar{\psi}\gamma\psi \right) =0,\\
        &\label{eq:Dirac bulk}e d_\omega \bar{\psi}\gamma^3 + \frac{1}{2}d_\omega e \bar{\psi}\gamma^3=0.
    \end{eqnarray}
\begin{remark}
    In the bulk, eq. (\ref{eq:torsion bulk}) is equivalent to $d_\omega e - \frac{1}{2}\bar{\psi}\gamma\psi=0$\footnote{That is because $e\wedge\cdot$ is an injective map when acting on $\Omega^{2}(M,\mathcal{V})$.}, implying that the background connection has torsion, while eq. \ref{eq:Dirac bulk} is equivalent to its complex conjugate, and can be re-interpreted (after imposing \eqref{eq:torsion bulk}) as the Rarita-Schwinger equation for a massless Majorana spinor in a curved background
        \begin{equation*}
            e\gamma^3 d_\omega \psi - \frac{1}{4}(\bar{\psi}\gamma\psi )\gamma^3 \psi=e\gamma^3 d_\omega \psi=0.
        \end{equation*}  
\end{remark}

\subsubsection{On-shell vs off-shell supersymmetry invariance}
So far we have been considering the connection as a dynamical field, in what is called the Palatini-Cartan formalism, also known as the first order formulation of (super)gravity, referring to the fact that only first order derivatives appear in the Lagrangian. If we impose \eqref{eq:torsion bulk}  in the absence of the gravitino, we obtain the torsionless condition, which, coupled with the metricity condition, gives the Levi-Civita connection as the pullback of omega by e. Upon application of this constraints one obtains the Einstein--Hilbert Lagrangian, which describes the second order formulation of gravity.

In the case of supergravity, \eqref{eq:torsion bulk} implies the non-vanishing of torsion, which will be quadratically dependent on the Majorana field $\psi$. Historically, the formulation of supergravity has been performed in the second--order formalism (the so called 'half-shell' case), i.e. after imposing the kinematical constraint \eqref{eq:torsion bulk}. 

In this setting, introducing a spinorial gauge parameter $\chi=\chi(x)$, defined to be an even\footnote{The reason we consider an unphysical Grassmann even fermion will be clear in the following section, as it will represent the ghost field associated to the gravitino} section of the Majorana spinor bundle, the infinitesimal supersymmetry transformations on the fields read
    \begin{align*}
        \delta_\chi e = - \bar{\chi}\gamma \psi,\qquad \delta_\chi \psi = d_\omega \chi,
    \end{align*}
with no need of specifying the variation of $\omega$ as it is constrained and can be obtained as a function of $\psi$ and $e$ from \eqref{eq:torsion bulk}.
It is indeed very quick to check the invariance of the action under these transformations
    \begin{align*}
        \delta\chi S_{SG} &=\int_\Sigma -i e \bar{\chi}\gamma \psi F_\omega + \frac{1}{3!} (-i\bar{\chi}\gamma\psi) (\bar{\psi}\gamma^3 d_\omega\psi) + \frac{1}{3!} e  (d_\omega \bar{\chi}\gamma^3 d_\omega \psi + \bar{\psi} \gamma^3[F_\omega,\chi] )\\
        &=\int_\Sigma -i e \bar{\chi}\gamma \psi F_\omega + \frac{1}{3} (e \bar{\psi}\gamma^3 [F_\omega,\chi] - [F_\omega,\bar{\psi}]\gamma^3 \chi])- \frac{1}{3!} \left( d_\omega e - \frac{1}{2}\bar{\psi}\gamma\psi \right) \bar{\chi}\gamma^3 d_\omega \psi =0,
    \end{align*}
having used the constraint \eqref{eq:torsion bulk}, identity \eqref{id: action of omega}, integration by parts, the Bianchi identity $d_\omega d_\omega (\cdot) = [F_\omega,\cdot]$ and the Fierz identity \eqref{Fierz:2} together with the flip relation \eqref{flip:3} to show $\bar{\psi}\gamma^3 d_\omega \psi \bar{\chi}\gamma \psi=d_\omega \bar{\psi}\gamma^3 \psi \bar{\chi}\gamma \psi = - d_\omega \bar{\psi}\gamma^3 \chi \bar{\psi}\gamma \psi - d_\omega \bar{\psi}\gamma^3 \psi \bar{\chi}\gamma \psi $, implying $\bar{\psi}\gamma^3 d_\omega \psi \bar{\chi}\gamma \psi= - \frac{1}{2}\bar{\psi}\gamma \psi \bar{\chi}\gamma^3 d_\omega\psi. $

If instead one keeps $\omega$ unconstrained, it is necessary to introduce the corresponding local SUSY transformation, which can be either derived by the requirement that the action remains invariant under the local supersymmetry (postulating the same transformations for $e$ and $\psi$), or by the analysis of the symplectic structure of the fields on the boundary $\Sigma=\partial M$, as it will be cleared in a future paper \cite{CFR25b}. We use here the first method, discarding the vanishing terms from the previous computations
    \begin{align*}
        \delta\chi S_{SG}&=\int_M -\frac{e^2}{2}d_\omega (\delta_\chi\omega ) -\frac{1}{3!}e \bar{\psi}\gamma^3[\delta_\chi \omega,\psi] - \frac{1}{3!} \left( d_\omega e - \frac{1}{2}\bar{\psi}\gamma\psi \right) \bar{\chi}\gamma^3 d_\omega \psi \\
        &=\int_M\left( d_\omega e -\frac{1}{2}\bar{\psi}\gamma\psi \right)\left( e \delta_\chi\omega - \frac{1}{3!}\bar{\chi}\gamma^3 d_\omega \psi\right), 
    \end{align*}
from which we obtain
    \begin{align}
        &\label{eq: local susy e}\delta_\chi e = - \bar{\chi}\gamma\psi,\\
        &\label{eq:local susy omega}e  \delta_\chi \omega = \frac{1}{3!}\bar{\chi}\gamma^3 d_\omega \psi,\\
        &\label{eq: local susy psi}\delta_\chi \psi = d_\omega \chi
    \end{align}
Notice that we have given only $e\delta_\chi \omega$ and not the explicit expression of $\delta_\chi \omega$ because it is not strictly necessary, since we are sure that $e\delta_\chi \omega$ uniquely determines the expression for $\delta_\chi \omega$. Indeed it suffices to notice that $W_e^{(1,2)}:= e\wedge \cdot\colon\Omega^1(M,\wedge^2 \mathcal{V})\rightarrow\Omega^2(M,\wedge^3\mathcal{V})$  provides an isomorphism,\footnote{A proof of this statement is found in \cite{Canepa:2024rib}} hence $\delta_\chi \omega$ is uniquely defined by the above equation.

In the following, it is convenient to adopt the following notation, setting $\Sigma\coloneqq\partial M$,
        \begin{equation*}
            \Omega^{(k,l)}:=\Omega^k(M,\wedge^l \mathcal{V}) \qquad \Omega_\partial^{(k,l)}:=\Omega^k(\Sigma,\wedge^l \mathcal{V}),
        \end{equation*}
furthermore, we define the coframes $e$ as those elements in $\Omega^{1,1}$ non-degenerate, hence $e\in\Omega_{\mathrm{n.d.}}^{(1,1)}$.

\section{The BV/BFV formalism in field theories}\label{sec: BV formalism}

The BV formalism was introduced by Batalin and Vilkovisky \cite{BV1, BV2, BV3} and later refined by Cattaneo, Mnev and Reshetikin \cite{CMR2,CMR2012,CMR2012b} to treat perturbative quantization of gauge field theories, possibly on manifolds with boundary. The main construction requires the space of fields to be treated as a $\mathbb{Z}$-graded\footnote{The grading is commonly referred to as "ghost degree", but here we consider for simplicity the total grading, i.e. the sum of all the degrees of a field belonging to various graded vector fields.} supermanifold, and to be endowed with a symplectic form and a cohomological Hamiltonian vector field encoding the classical symmetries of the system.

Typically, a field theory on $M$ is the assignment of a space of fields $F_M$, usually defined locally to be the sections of a vector bundle $F\rightarrow M$, and of an action functional $S_M$ on $F_M$, whose variation produces the equations of motion of the theory, also known as Euler-Lagrange equations. The symmetries of the theory are defined by vector fields on $F_M$  leaving the action invariant, i.e. $X\in\mathfrak{X}(F_M)$ such that $L_X(S_M)=0$, where $L_X$ is the Lie derivative on the space of fields.

In order to have a clear definition of the objects above, one needs to take into consideration the variational calculus\footnote{To be precise, consider the infinite jet bundle $J^\infty F$ and the sections on it $\Gamma(M,J^\infty F)$. It is a well known fact that there exists a bicomplex \cite{anderson, Zuckerman} defined by differential forms on $\Gamma(M,J^\infty F)$, in particular the de-Rham differential splits into a horizontal and a vertical differental $d_\infty=d_H+d_V$ satisfying $d_H^2=d_V^2=d_H d_V + d_V d_H = 0$, defining a double degree on $\Omega^n(J^\infty F)=\bigoplus_{p+q=n}\Omega^{p,q}(J^\infty F)$,where $p$ is the horizontal degree and $q$ the vertical one.

In order to obtain a well defined local calculus, we consider the infinite jet prolongation $j^\infty:F_M:=\Gamma(M, F) \rightarrow \Gamma(M,J^\infty F)$ and precompose it with the evaluation map $ev:M\times F_M\rightarrow F\colon(x,\phi)\mapsto \phi(x) $ to obtain 
\begin{align*}
    e_\infty \colon {M\times F_M} \xrightarrow{(\mathrm{id},j^\infty)} {M\times \Gamma(M,J^\infty F)} \xrightarrow{\mathrm{ev}} {J^\infty F}
\end{align*}
and define local forms on $M\times F_M$ by pulling back forms on $J^\infty F$ along $e_\infty$, i.e. $\Omega^{p,q}_{loc}(M\times F_M):=e_\infty^*\Omega^{p,q}(J^\infty F)$. $\Omega^{p,q}_{loc}(M\times F_M)$ is then also endowed with a double degree and two differentials $d$ and $\delta$, respectively the de-Rham differential on differential forms on $M$ (which take the role of horizontal forms in the bicomplex) and the vertical differential $\delta$, which encodes the variation of a functional on $M\times F_M$ when the field configuration is varied. In this setting, one defines the Lagrangian $L_M$ to be a local $(D,0)$-form (setting $D=$dim$M$) and the action as in integrated 0-form on $F_M$, defined by $S_M=\int_M L_M$.
} on $F_M$. In particular, $F_M$ is an infinite-dimensional manifold, inheriting the structure of a Fréchet space, and as such it might be complicated to generalize classical objects defined in finite dimensions. Nevertheless, for the problem at hand, we only need to assume Cartan calculus to be defined on it.

\begin{definition}
    A BV manifold on $M$ is the assignment of data $(\mathcal{F}_M,\mathcal{S}_M,Q,\varpi_M)$, where $(\mathcal{F}_M,\varpi_M)$ is a $\mathbb{Z}$-graded manifold endowed with a -1-symplectic form $\varpi_M$,  and $\mathcal{S}_M$ and $Q$ are respectively a degree 0 funcional (called BV action) and a degree 1 vector field on $\mathcal{F}_M$ such that 
    \begin{itemize}
        \item $\iota_{Q}\varpi_M=\delta\mathcal{S}_M$, i.e. $Q$ is the Hamiltonian vector field of $\mathcal{S}_M$;
        \item $Q^2=\frac{1}{2}[Q,Q]=0$, i.e. $Q$ is cohomological.
    \end{itemize}
\end{definition}

\begin{remark}
    As a consequence of $Q$ being cohomological, the BV action satisfies the classical master equation
    \begin{equation}
        (S,S)=0.
    \end{equation}
\end{remark}

In the context of field theory, $\mathcal{F}_M$ is a graded manifold whose body is given by the classical space of field $F_M$, while the graded part contains  the ghosts (related to the symmetries of the theory), the anti-fields. $\mathcal{S}_M$ is an extension of the classical action, to which it reduces on the body, containing terms depending on all the other fields, in such a way that its Hamiltonian vector field $Q$ encodes all the symmetries of the classical theory and is cohomological.

In good cases, such symmetries form a distribution $D\subset \mathfrak{X}(F_M)$, which might be the action of a Lie algebra of a certain Lie group, in which case the BV formalism reduces to BRST (see \cite{Mnev2017}), but in general one only has that the distribution is involutive on the Euler-Lagrange locus $EL_M:=\{ \phi\in F_M \hspace{1mm}\vert \hspace{1mm} \delta S \vert_\phi=0 \} $. 
At this point, the simplest BV manifold one can construct is $\mathcal{F}_M:=T^*[-1]D[1] $, where the vector fields encoding the symmetries are now promoted to fields of degree one (the ghosts) and the odd cotangent fibers define the anti-fields of degree -1 and -2. Such graded spaces of fields can now be endowed with the canonical -1--symplectic form defined on a -1--shifted cotangent bundle. In this setting, denoting by $\Phi=(\Phi^\alpha)$ a multiplet (containing fields and ghosts) in $D[1]$ and by $\Phi^\dagger=(\Phi^\dagger_\alpha)$ its canonical conjugate (containing  the anti-fields) in the fiber of $T^*[-1]D[1]$, and letting $Q_0$ be the vector field encoding the classical symmetries of the theory, one can define the BV action  as a linear functional in the anti-fields(ghosts)
    \begin{align*}
        &\varpi_M = \int_M \delta \Phi^\dag_\alpha \delta \Phi^\alpha,\\
        &\mathcal{S}_M=S_M +\int_M  \Phi^\dag_\alpha Q_0(\phi^\alpha),\\
    \end{align*}
obtaining
    \begin{align*}
        Q(\Phi^\alpha)=Q_0(\Phi^\alpha), \qquad \mathrm{and} \qquad Q(\Phi^\dag_\alpha)= \frac{\delta L_M}{\delta \Phi^\alpha} - (-1)^\beta \Phi^\dag_\beta \frac{\delta (Q_0 \phi^\beta)}{\delta \Phi^\alpha},
    \end{align*}
where $L_M$ is the Lagrangian density and $\alpha:=|\Phi^\alpha| $ is the parity of $\Phi^\alpha$. In the general case, the BV action starts like in the BRST case but has further terms, nonlinear in the antifields.
\begin{remark}
Notice that $Q$ on the anti-fields contains a term $\frac{\delta L_M}{\delta \Phi^\alpha}$ which, on the body, defines the equations of motion related to the field $\Phi^\alpha$. Therefore, when computing the degree--0 cohomology of $Q$, one can intuitively see how this is related to the gauge-invariant functions on the Euler-Lagrange locus, as they are given by 
        \begin{equation*}
            \frac{\ker{Q:C^0\rightarrow C^1}}{\text{Im}Q:C^{-1}\rightarrow C^0 } \simeq C^\infty\left\{ \Phi^\alpha \hspace{1mm}\big\vert\hspace{1mm} \frac{\delta L_M}{\delta \Phi^\alpha}=0 \right\}/\{ \text{gauge transformation} \},
        \end{equation*}
    where $C^k$ are the functions of ghost degree $k$ on $\mathcal{F}_M$.   
\end{remark}

The case of this paper is not of BRST type.\footnote{$Q^2$ will vanish only on shell, i.e. it will contain terms proportional to the equations of motion.} However, the BV action stops at the next order, quadratic in the antifields; i.e., it has the form 
    \begin{align*}
        &\mathcal{S}_M=S_M +\int_M  \Phi^\dag_\alpha Q_0(\phi^\alpha) + \frac{1}{2}\Phi^\dag_\alpha \Phi^\dag_\beta M^{\alpha\beta}(\Phi) , \\ 
    \end{align*}
which modifies the Hamiltonian vector field as 
    \begin{align*}
        &Q(\Phi^\alpha)=Q_0(\Phi^\alpha) + \Phi^\dag_\beta M^{\alpha\beta} ,\\ 
        &Q(\Phi^\dag_\alpha)= \frac{\delta L_M}{\delta \Phi^\alpha} - (-1)^\beta \Phi^\dag_\beta \frac{\delta (Q_0 \phi^\beta)}{\delta \Phi^\alpha} + \frac{(-1)^{\beta+\gamma} }{2 }  \Phi^\dag_\beta  \Phi^\dag_\gamma \frac{\delta M^{\beta\gamma} }{\delta \Phi^\alpha}  .
    \end{align*}
In this case, the classical master equation reads    
    \begin{align*}
        (\mathcal{S}_M,\mathcal{S}_M)&=\int_M Q_0(L_M) - (-1)^\beta \Phi^\dag_\beta \left(  Q_0^2 \Phi^\alpha - (-1)^{\beta(\alpha + 1)}\frac{\delta L_M}{\delta \Phi^\alpha} M^{\alpha\beta}   \right)\\
        &\qquad + \frac{(-1)^{\beta+\gamma}}{2}\Phi^\dag_\gamma \Phi^\dag_\beta \left(  Q_0(M^{\beta\gamma} ) -(-1)^{\gamma + \beta\alpha} \frac{\delta Q_0 \Phi^\beta}{\delta \Phi^\alpha} M^{\alpha\gamma} -(-1)^{\beta+ \gamma\alpha} \frac{\delta Q_0 \Phi^\gamma}{\delta \Phi^\alpha} M^{\alpha\beta} \right)  \\
        & \qquad +\frac{(-1)^{\alpha\beta}}{2}\Phi^\dag_\beta \Phi^\dag_\rho \Phi^\dag_\gamma \frac{\delta M^{\rho\gamma} }{\delta \Phi^\alpha} M^{\alpha\beta}.
    \end{align*}
We know that $Q_0(L_M)=0$\footnote{Under certain assumptions it could also be a boundary term.}  by definition of $Q_0$, while the remaining terms at each order in the anti-fields(ghost) must be imposed separately,
    \begin{eqnarray}
        &\label{eq: def Mab} Q_0^2 \Phi^\alpha - (-1)^{\beta(\alpha + 1)}\frac{\delta L_M}{\delta \Phi^\alpha} M^{\alpha\beta} =0, \\
        &\label{eq: consist SS 1}Q_0(M^{\beta\gamma} ) -(-1)^{\gamma + \beta\alpha} \frac{\delta Q_0 \Phi^\beta}{\delta \Phi^\alpha} M^{\alpha\gamma} -(-1)^{\beta+ \gamma\alpha} \frac{\delta Q_0 \Phi^\gamma}{\delta \Phi^\alpha} M^{\alpha\beta} =0,\\
        &\label{eq: consist SS 2}\frac{\delta M^{\rho\gamma} }{\delta \Phi^\alpha} M^{\alpha\beta} = 0.
    \end{eqnarray}
One can then use \eqref{eq: def Mab} to fix $M^{\alpha\beta} $ and then check that \eqref{eq: consist SS 1} and \eqref{eq: consist SS 2} hold.
\begin{remark}
    It is just a matter of computations to show that $Q^2(\Phi^\alpha)=0$ implies \eqref{eq: def Mab}, \eqref{eq: consist SS 1} and \eqref{eq: consist SS 2}. Hence it is not needed to show $Q^2(\Phi^\dag_\alpha)=0$, as it follows naturally.
\end{remark}

\subsection{The case of gravity in the PC formalism}\label{subsec: BV PC}
Palatini-Cartan gravity is recovered from supergravity by "turning off" the gravitino interaction, and was studied within the BV formalism in \cite{Piguet:2000fy} and later refined in \cite{CS2017}. The field content is just that of the vielbein and the spin connection, while the symmetries are defined by the diffeomorphisms and the internal (i.e. with respect to the Minkwoski bundle indices) gauge transformations, amounting to the $SO(3,1)$ invariance. 

\begin{theorem}[\cite{CS2017}]
    The collection $(\mathcal{F}_{PC},\varpi_{PC},Q_{PC},\mathcal{S}_{PC})$ defines a BV structure, where $\mathcal{F}_{PC}:=T^*[-1]F_{PC} $ and 
    \begin{equation*}
        {F}_{PC}=\Omega_{\mathrm{n.d.}}^{(1,1)}\oplus \mathcal{A}_M  \oplus \Omega^{(0,2)}[1] \oplus \mathfrak{X}[1](M) \ni (e,\omega,c,\xi).
    \end{equation*}
    The symplectic form is canonically defined as
        \begin{equation*}
            \varpi_{PC}=\int_M \delta e \delta e^\dag + \delta \omega \delta \omega^\dag + \delta c \delta c^\dag + \iota_{\delta \xi}\delta \xi^\dag,
        \end{equation*}
    while the BV action reads
    \begin{align*}
        \mathcal{S}_{PC} &=\int_M \frac{e^2}{2}F_\omega   -(L_\xi^\omega e - [c,e] ) e^\dag + (\iota_\xi F_\omega - d_\omega c )\omega^\dag  \\
        &\qquad  + \frac{1}{2}(\iota_\xi\iota_\xi F_\omega - [c,c] )c^\dag +\frac{1}{2}\iota_{[\xi,\xi]} \xi^\dag.
    \end{align*}

    Lastly, one easily recovers the cohomological vector field acting on the fields and the ghosts as
    \begin{align*}
        & Q_{PC} e = L_\xi^\omega e - [c,e] & Q_{PC} \omega = \iota_\xi F_\omega - d_\omega c \\
        &Q_{PC} c = \frac{1}{2}( \iota_\xi\iota_\xi F_\omega -[c.c] ) & Q_{PC} \xi = \frac{1}{2}[\xi,\xi] .
    \end{align*}
\end{theorem}
\begin{remark}
    One can distinguish the diffeomorphism symmetry $\delta_\xi$ generated by an odd vector field $\xi\in\mathfrak{X}[1](M)$ and the internal gauge symmetry $\delta_c$ generated by an odd section of the Lie algebra of the Lorentz group $c\in\Gamma(\wedge^2 \mathcal{V})=\Omega^{(0,2)}$. 
\end{remark}

\section{The $\mathcal{N}=1$, $D=4$ Supergravity BV action in the first order formalism}\label{sec: BV Sugra}

A BV description of on-shell $\mathcal{N}=1,$ $D=4$ supergravity has been provided in \cite{Baulieu:1990uv}, where it was shown that the BV action is of rank 2 (i.e. quadratic in the anti-fields). However, to the best of our knowledge, no off-shell BV description of it has been obtained.

We start here by applying the simplest procedure from section \ref{sec: BV formalism}, defining the space of BV fields as 
    \begin{equation*}
        \mathcal{F}_{SG}=T^*[-1]\big(\Omega^{(1,1)}_{\mathrm{n.d.}} \otimes \mathcal{A}_M \otimes \Omega^{1}(M,\Pi \mathbb{S}_M) \otimes \Omega^{(0,2)}[1] \times \mathfrak{X}[1](M) \otimes \Gamma[1](M,\Pi \mathbb{S}_M) \big),
    \end{equation*}
where
    \begin{itemize}
        \item $e\in\Omega^{(1,1)}_{\mathrm{n.d.}}, $ $\omega\in\mathcal{A}_M$ and $\psi\in\Omega^{1}(M,\Pi \mathbb{S}_M)$ are the classical fields;
        \item $c\in \Omega^{(0,2)}[1]=\Gamma[1](M,\wedge^2 \mathcal{V})\simeq\Gamma[1](M, \mathfrak{so}(1,3))$, $\xi \in \mathfrak{X}[1](M) $ and $\chi \in \Gamma[1](M,\Pi \mathbb{S}_M) $ are the ghost fields,\footnote{Note that all the ghosts have ghost number 1, yet $\chi$, unlike $c$ and $\xi$, has even Grassmann parity.} seen as odd generators respectively to the internal Lorentz symmetry, the diffeomorphism symmetry and the local supersymmetry;
        \item $e^\dagger \in \Omega^{(3,3)}[-1] $, $\omega^\dag \in \Omega^{(3,2)}[-1] $ and $\psi_\dag\in\Omega^{(3,4)}[-1](M,\Pi \mathbb{S}_M)$ are the field momenta, while $c^\dag \in\Omega^{(4,2)}[-2] $, $\xi^\dag\in \Omega^1(M)[-2]\otimes\Omega^{(4,4)} $ and $\chi_\dag\in\Omega^{(4,4)}[-2](M,\Pi \mathbb{S}_M) $ are the ghost momenta.
    \end{itemize} 
The -1--symplectic forms reads
        \begin{equation}
            \varpi_{SG}=\int_M \delta e \delta e^\dag + \delta \omega \delta\omega^\dag + i \delta \bar{\psi}\delta \psi_\dag + \delta c \delta c^\dag + \iota_{\delta \xi}\delta\xi^\dag + i \delta\bar{\chi}\delta \chi_\dag.
        \end{equation}

Our first attempt of finding a suitable BV action requires finding the vector field $Q_0$ describing the symmetries of the theory. We define\footnote{One could obtain the correct SUSY transformations by inspecting the boundary structure and phase space Hamiltonian of supergravity.} 
    \begin{align*}
        & Q_0 e = L_\xi^\omega e - [c,e] +  \bar{\chi}\gamma\psi & Q_0 \omega = \iota_\xi F_\omega - d_\omega c + \delta_\chi \omega \\
        &Q_0 \psi= L_\xi^\omega \psi - [c,\psi] - d_\omega \chi & Q_0 \xi = \frac{1}{2}[\xi,\xi] + \frac{1}{2}\varphi \\
        & Q_0 c = \frac{1}{2}( \iota_\xi\iota_\xi F_\omega -[c.c] ) + \iota_\xi \delta_\chi\omega  & Q_0 \chi = L_\xi^\omega \chi - [c,\chi] - \frac{1}{2}\iota_\varphi \psi,
    \end{align*}
where $e\delta_\chi\omega =- \frac{1}{3!}\bar{\chi}\gamma^3 d_\omega\psi$ and $\varphi^\mu= \bar{\chi}\gamma^\mu\chi$. In particular, for the fields on which it is defined, one can notice that $Q_0=Q_{PC} + \delta_\chi$, having borrowed $Q_{PC}$ from \cite{CS2017}. Since we know $Q_{PC}^2 =0$, we obtain $$Q_0^2=[Q_{PC},\delta_\chi] + \delta_\chi^2.$$

The classical action $\mathcal{S}_0$ is then complemented with a contribution $s_1$ linear in the anti--fields, obtaining
    \begin{align*}
        \mathcal{S}_1&=\mathcal{S}_0+s_1=\int_M \frac{e^2}{2}F_\omega + \frac{1}{3!} e \bar{\psi}\gamma^3 d_\omega\psi \\
        &\quad+\int_M -(L_\xi^\omega e - [c,e] + \bar{\chi}\gamma\psi) e^\dag + (\iota_\xi F_\omega - d_\omega c + \delta_\chi \omega)\omega^\dag  \\
        &\qquad -i (L_\xi^\omega \bar{\psi} - [c,\bar{\psi}] -d_\omega \bar{\chi})\psi_\dag + \left(\frac{1}{2}\iota_\xi\iota_\xi F_\omega -\frac{1}{2} [c,c] + \iota_\xi \delta_\chi\omega  \right)c^\dag   \\
        &\qquad +\frac{1}{2}(\iota_{[\xi,\xi]} + \iota_\varphi)\xi^\dag  - i\left(L_\xi^\omega\bar{\chi} - [c,\bar{\chi}]  -\frac{1}{2}\iota_\varphi \bar{\psi} \right)\chi_\dag.
    \end{align*}

In principle, to check the classical master equation $\{\mathcal{S}_1,\mathcal{S}_1\}_{BV}=0$ it is sufficient to prove $Q_0^2=0$ on the fields and ghost. Proceeding by stages, we first obtain
    \begin{align*}
        & \delta_\chi^2 e = -\frac{1}{2}\mathrm{L}_\varphi^\omega e + \frac{1}{2}\iota_\varphi\left( d_\omega e - \frac{1}{2}\bar{\psi}\gamma\psi \right)\\
        & \delta_\chi^2 \psi=- \frac{1}{2}\mathrm{L}_\varphi^\omega \psi + \frac{1}{2}\vf d_\omega\psi -  \left(\bar{\chi} \kappa(<\bar{e},\ugam d_\omega \psi>) + \frac{1}{8} \bar{\chi} \iota_{\hat{\gamma}}\iota_{\hat{\gamma}}(\ugam d_\omega\psi) \right)\chi \\
        & e \delta_\chi^2 \omega = -\frac{1}{2}e\vf F_\omega +\frac{1}{2}\vf\left(  eF_\omega+ \frac{1}{3!}\bar{\psi}\gamma^3 d_\omega \psi \right) - \frac{1}{2\cdot 3!}\bar{\psi}\vf(\gamma^3 d_\omega\psi)\\
        &\hspace{11.5mm}-\frac{1}{3!}\bar{\psi}\gamma^3\chi\left(\bar{\chi} \kappa(<\bar{e},\ugam d_\omega \psi>) + \frac{1}{8} \bar{\chi} \iota_{\hat{\gamma}}\iota_{\hat{\gamma}}(\ugam d_\omega\psi) \right)\\
        &   \delta^2_\chi c= \frac{1}{2}\iota_\varphi \delta_\chi \omega + \iota_\xi \delta_\chi^2\omega \qquad \qquad \qquad \quad \delta_\chi^2 \chi= -\frac{1}{2}\mathrm{L}^\omega_\varphi\chi & \delta^2_\chi \xi=0,
    \end{align*}
where $\hat{\gamma}:=\gamma^\mu\partial_\mu=e^\mu_a \gamma^a \partial_\mu$ and the map $<e,->$ is defined via the inverse vielbein as 
    \begin{align*}
        <e,->\colon\Omega^{(i,j)}&\longrightarrow\Omega^{(i-1,j+1)}\\
        \sigma&\longmapsto v_a\eta^{ab}e^\mu_b\iota_{\partial_\mu}\sigma.
    \end{align*}
Notice that, as expected, the square of the supersymmetry transformation is proportional to the diffeomorphisms\footnote{This is in line with the fact that supersymmetry squares to the translations, which in their local version are realized by the diffeomorphisms.}  with respect to the generator $\varphi:=\bar{\chi}\hat{\gamma} \chi$, plus a term which is proportional to the equations of motion. The full computation of $Q_0^2$ is found in \ref{app: Q_0^2}, it gives us
    \begin{align*}
        & Q_0^2 e = \frac{1}{2}\iota_\varphi\left( d_\omega e - \frac{1}{2}\bar{\psi}\gamma\psi \right) \\
        & Q_0^2 \psi=\frac{1}{2}\iota_\varphi d_\omega \psi  -  \left(\bar{\chi} \kappa(<\bar{e},\ugam d_\omega \psi>) + \frac{1}{8} \bar{\chi} \iota_{\hat{\gamma}}\iota_{\hat{\gamma}}(\ugam d_\omega\psi) \right)\chi \\
        & e Q_0^2 \omega =\frac{1}{2}\iota_\varphi\left(e F_\omega +\frac{1}{3!}\bar{\psi}\gamma^3 d_\omega\psi  \right) + \frac{1}{2\cdot 3!}\bar{\psi}\gamma^3 \iota_\varphi d_\omega \psi\\
        &\hspace{11.5mm}-\frac{1}{3!}\bar{\psi}\gamma^3\chi\left(\bar{\chi} \kappa(<\bar{e},\ugam d_\omega \psi>) + \frac{1}{8} \bar{\chi} \iota_{\hat{\gamma}}\iota_{\hat{\gamma}}(\ugam d_\omega\psi) \right)\\
        &Q_0^2 c= \frac{1}{2}\iota_\varphi \delta_\chi \omega + \iota_\xi Q_0^2\omega \qquad \qquad \qquad \quad  Q_0^2 \chi= 0 & Q_0^2 \xi=0,
    \end{align*}
This tells us that the BV description of $\mathcal{N}=1$, $D=4$ SuGra is at least of second rank, hence we need to correct the action. 

\subsection{The second rank BV action}

Before continuing, for computational purposes, it is convenient to redefine some of the fields. In particular, using \ref{lem: useful isos}.\ref{lem: iso W2 (2,0)} and looking at the diagram \ref{diag: prop e bulk}, we notice that one can uniquely define $\check{c}\in\Omega^{(2,0)}[-1] $ and $\check{\omega}\in\Omega^{(2,1)}[-1] $ such that 
\begin{equation}
    c^\dag= \frac{e^2}{2} \check{c} \qquad \text{and}\qquad \omega^\dag= e \check{\omega}. 
\end{equation}
With this redefinition, we then see
    \begin{align*}
         \frac{e^2}{2}Q_0^2 c=& \frac{i}{8}\bar{\chi}\iota_\varphi(\mathrm{EoM}_\psi)-\frac{1}{8\cdot 3!}\iota_\varphi\big( (\mathrm{EoM}_\omega)\bar{\chi}\gamma^3\psi \big  )-\frac{1}{2}\iota_\xi e \iota_\varphi(\mathrm{EoM}_e)\\
         & -\frac{1}{2\cdot 3! }\iota_\xi e \bar{\psi}\gamma^3\iota_\varphi d_\omega \psi+ \iota_\xi \left( \frac{e}{4} \iota_\varphi (\mathrm{EoM}_e) \right) + \frac{i}{8}\iota_\xi \iota_\varphi(\bar{\psi} \mathrm{EoM}_\psi)\\
         & -\frac{e}{2}\iota_\xi\left(\frac{1}{3!}\bar{\psi}\gamma^3\chi\left(\bar{\chi} \kappa(<\bar{e},\ugam d_\omega \psi>) + \frac{1}{8} \bar{\chi} \iota_{\hat{\gamma}}\iota_{\hat{\gamma}}(\ugam d_\omega\psi) \right)\right)\\
         &+\frac{1}{2}\iota_\xi e\frac{1}{3!}\bar{\psi}\gamma^3\chi\left(\bar{\chi} \kappa(<\bar{e},\ugam d_\omega \psi>) + \frac{1}{8} \bar{\chi} \iota_{\hat{\gamma}}\iota_{\hat{\gamma}}(\ugam d_\omega\psi) \right) 
    \end{align*}


There are still some terms that are not immediately recognizable as proportional to the equations of motion. In order to achieve that, one needs lemmata \ref{lem: iso (1,0) (3,4)}, \ref{lem: splitting (3,1)} and \ref{lem: splitting (2,1)}. In particular, setting $\ugam:=[e,\gamma]= \gamma_\mu dx^\mu$, thanks to \ref{lem: iso (1,0) (3,4)} we can redefine $\psi_\dag$ as 
    \begin{equation*}
        \psi_\dag:= \frac{1}{3!}e \gamma^3 \underline{\gamma} \psi^0_\dag, 
    \end{equation*}
while from  \ref{lem: splitting (3,1)} and \ref{lem: splitting (2,1)} we have the following maps
        \begin{align*}
            &\alpha\colon\Omega^{(3,1)}(\Pi \mathbb{S}_M)\rightarrow \Omega^{(1,0)}(\Pi \mathbb{S}_M) &\beta\colon\Omega^{(3,1)}(\Pi \mathbb{S}_M)\rightarrow\ker{(\gamma^3_{(3,1)})}\\
            &\kappa\colon\Omega^{(2,1)}(\Pi \mathbb{S}_M)\rightarrow\Omega^{(1,0)}(\Pi \mathbb{S}_M) & \varkappa\colon \Omega^{(2,1)}(\Pi \mathbb{S}_M)\rightarrow\ker{(\ugam\gamma^3_{(2,1)})}
        \end{align*}
        such that for all $\theta\in\Omega^{(3,1)} $ and $\omega\in\Omega^{(2,1)}$ one has
            \begin{align*}
               & \theta= i e \ugam\alpha(\theta) + \beta(\theta), \qquad\omega = e \kappa(\omega) + \varkappa(\omega).
            \end{align*}
Lastly, one can use the fact that $v_a v_b v_c v_d = \epsilon_{abcd}\mathrm{Vol}_V$ and \eqref{id: gamma5gammac} to show that the equation of motion for the gravitino reduces to 
    \begin{equation*}
        \frac{i}{3}\left( e\gamma^3 d_\omega \psi -\frac{1}{2}d_\omega e \gamma^3 \psi \right)= -\frac{1}{3}\gamma^5\left(\ugam d_\omega\psi - \frac{1}{2}[d_\omega e,\gamma]\psi \right)\mathrm{Vol}_V=0.
    \end{equation*}
In the end, from the terms of the kind $\int \Phi^\dag_\alpha Q_0^2\Phi^\alpha $ inside $(\mathcal{S}_1,\mathcal{S}_1)$, we can use \eqref{eq: def Mab} to obtain the coefficients of the rank--2 action, obtaining $\mathcal{S}_2=\mathcal{S}_0+s_1+s_2$, with
    \begin{align}\label{eq: quadratic BV action}
        \nonumber s_2&=\int_M \frac{1}{2}\left(\check{\omega}-\frac{1}{2}e\iota_\xi \check{c} - \check{c}\iota_\xi e\right)\vf e^\dag +\frac{1}{4}\left( \frac{1}{2}\bar{\psi}^0_\dag\ugam + \alpha(\check{\omega}\bar{\psi})\ugam -\frac{i}{2}\iota_\xi\check{c}\bar{\psi}-\alpha(\check{c}\iota_\xi e \bar{\psi})\ugam -\frac{i}{2}\check{c}\bar{\chi}\right)\vf\psi_\dag\\
        \nonumber&\qquad + \frac{i}{4\cdot 3!}\left(\frac{1}{2}\alpha(\check{\omega}\bar{\psi})\ugam -\frac{i}{2}\iota_\xi\check{c}\bar{\psi}-\alpha(\check{c}\iota_\xi e \bar{\psi})\ugam -\frac{i}{2}\check{c}\bar{\chi}\right)\gamma^3\vf(\check{\omega}\psi)\\
        &\qquad -\frac{i}{2\cdot 3!}\left( \frac{1}{2}\bar{\psi}^0_\dag\ugam+\frac{1}{2}\alpha(\check{\omega}\bar{\psi})\ugam -\frac{i}{2}\iota_\xi\check{c}\bar{\psi}-\alpha(\check{c}\iota_\xi e \bar{\psi})\ugam \right)\gamma^3\chi <e,\bar{\chi}[\check{\omega},\gamma]\psi>\\
        \nonumber&\qquad +\frac{1}{2\cdot 3!}\left( \frac{1}{4}\bar{\psi}^0_\dag\ugam-\frac{i}{2}\iota_\xi\check{c}\bar{\psi}-\alpha(\check{c}\iota_\xi e \bar{\psi})\ugam \right)\gamma^3\chi <e,\bar{\chi}\ugam^2\psi^0_\dag>\\
        \nonumber&\qquad -\frac{1}{32}\left(i\bar{\psi}_\dag \chi +\frac{1}{3!}(\check{\omega} - e \iota_\xi \check{c} - 2 \check{c}\iota_\xi e  )\bar{\psi}\gamma^3 \chi \right)\bar{\chi}\igam\igam([\check{\omega},\gamma]\psi)\\
        \nonumber&\qquad - \frac{i}{32}\left(i\bar{\psi}_\dag\chi + \frac{1}{3!}(e\iota_\xi \check{c} + 2\check{c}\iota_\xi e )\bar{\psi}\gamma^3\chi  \right)\bar{\chi}\igam\igam(\ugam^2\psi^0_\dag),
    \end{align}
Now, letting $\mathbb{q}$ be the Hamiltonian vector field of $s_2$, we obtain $Q=Q_0+\mathbb{q}$, and, after a long but straightforward computation.  
    \begin{align*}
         &\mathbb{q}_e =  \frac{1}{2}\iota_\varphi\check{\omega} -\frac{1}{2}\iota_\varphi\check{c}\iota_\xi e -\frac{1}{4}\iota_\varphi(e \iota_\xi \check{c} ) \\
         &e \mathbb{q}_\omega =   \frac{1}{2} \vf e^\dag +\frac{i}{4\cdot 3!} \vf(\bar{\psi^0_\dag}\ugam)\gamma^3 \psi +\frac{i}{4\cdot 3! }\bar{\psi}\gamma^3\vf\left(\ugam\alpha(\check{\omega}\psi )\right) - \frac{1}{8\cdot 3!} \vf \check{c} \bar{\chi}\gamma^3 \psi - \frac{1}{8\cdot 3!} \iota_\xi \check{c} \bar{\psi}\gamma^3 \vf\psi \\
         &\qquad \hspace{3mm}   -\frac{i}{4\cdot 3!} \bar{\psi}\gamma^3 \vf\left( \ugam\alpha(\check{c}\iota_\xi e \psi) \right)  + \frac{1}{2\cdot 3!}\bar{\psi}\gamma^3\chi\kappa\left[ <e, \bar{\chi}\left( -\frac{i}{2}\ugam^2\psi_0^\dag - [\check{\omega},\gamma]\psi - \frac{1}{2}\ugam\iota_\xi\check{c}\psi - \iota_\xi\ugam \check{c}\psi \right) > \right]\\
         &\qquad \hspace{3mm} + \frac{1}{16\cdot 3!}\bar{\psi}\gamma^3\chi \bar{\chi}\igam\igam\left(  -\frac{i}{2}\ugam^2\psi_0^\dag - [\check{\omega},\gamma]\psi - \frac{1}{2}\ugam\iota_\xi\check{c}\psi - \iota_\xi\ugam \check{c}\psi  \right)\\
        &\mathbb{q}_\psi = \frac{i}{4}\vf(\ugam \psi^0_\dag) -\frac{i}{4}\vf\left( \ugam\alpha( \check{\omega}\psi )  \right)   - \frac{i}{4}\vf\left( \ugam\alpha( \check{c}\iota_\xi e \psi )   \right) + \frac{1}{8}\vf\check{c} \chi - \frac{1}{8} \vf(\iota_\xi\check{c}\psi )\\
        & \hspace{7mm} +\frac{i}{4}\chi\kappa\left( <\bar{e}, \bar{\chi}\ugam^2\psi^0_\dag +i \bar{\chi}[\check{\omega}-\frac{1}{2}\iota_\xi \check{c} e - \iota_\xi e \check{c},\psi]> \right)+\frac{1}{16}\chi\bar{\chi}\igam\igam(\ugam^2\psi^0_\dag +i [\check{\omega}-\frac{1}{2}\iota_\xi \check{c} e - \iota_\xi e \check{c},\psi])\\
        &\frac{e^2}{2} \mathbb{q}_c = -\frac{i}{8}\bar{\chi}\vf\psi_\dag -\frac{i}{8\cdot 3!}\vf(\check{\omega }\bar{\chi}\gamma^3 \psi) -\frac{1}{2}\iota_\xi e \vf e^\dag+ \frac{1}{4}\iota_\xi( e \vf e^\dag)- \frac{i}{4\cdot 3!}\vf(\bar{\psi}^0_\dag \ugam) \gamma^3 \iota_\xi e \psi  \\
        &\qquad \hspace{3mm} + \frac{i}{4\cdot 3!}\vf\left(\alpha(\check{\omega}\bar{\psi}) \ugam\right) \gamma^3 \iota_\xi e \psi -\frac{i}{8}\iota_\xi(\bar{\psi} \vf\psi_\dag) - \frac{1}{8\cdot 3!}\iota_\xi(\check{\omega}\bar{\psi}\gamma^3 \vf\psi)\\
        &\qquad \hspace{3mm}+\frac{1}{4\cdot3!} \iota_\xi\left(\bar{\psi}\gamma^3\chi<e, \bar{\chi}([\check{\omega},\gamma]\psi + i\ugam^2\psi^0_\dag) >  \right)-\frac{1}{2\cdot 3!}\iota_\xi e \bar{\psi}\gamma^3\chi\kappa\left(<e, \bar{\chi}([\check{\omega},\gamma]\psi + i\ugam^2\psi^0_\dag) > \right)\\
        &\qquad \hspace{3mm} + \frac{1}{32\cdot 3!}\iota_\xi e \bar{\psi}\gamma^3\chi \bar{\chi}\igam\igam([\check{\omega},\gamma]\psi + i\ugam^2\psi^0_\dag) - \frac{1}{32\cdot3!} e \iota_\xi\left(\bar{\psi}\gamma^3\chi \bar{\chi}\igam\igam([\check{\omega},\gamma]\psi + i\ugam^2\psi^0_\dag)\right),
    \end{align*}
while one can immediately see $\mathbb{q}_\chi=0$ and $\mathbb{q}_\xi=0$.

Unfortunately, it turns out that \eqref{eq: quadratic BV action} is not yet the full rank--2 action. Indeed, as seen from computations in the appendix \ref{app: Q^2}, one needs to require the cohomological vector field $Q$ along $c$ to contain terms proportional to $\iota_\xi Q\omega$. This is not the case here as $s_2$ is missing terms quadratic in the antighost $\check{c}$.
\begin{remark}
    As  stated above, one can use equation \eqref{eq: def Mab} 
    \begin{equation*}
        Q_0^2 \Phi^\alpha - (-1)^{\beta(\alpha + 1)}\frac{\delta L_M}{\delta \Phi^\alpha} M^{\alpha\beta} =0
    \end{equation*}
    to define $M^{\alpha\beta}(\Phi)$, which are exactly the coefficients appearing in the quadratic part of the action, where the equation of motion $\frac{\delta L_M}{\delta \Phi^\alpha}$ is replaced by the corresponding antifield $\Phi^\dag_\alpha$. However, since there is no equation of motion for the ghosts, and in particular no equation of motion for $c$, the terms quadratic in $\check{c}$ have to be found by hand by checking $Q^2=0$, or equivalently by imposing the consistency equations \eqref{eq: consist SS 1},\eqref{eq: consist SS 2}. 
\end{remark}
As it turns out, defining $e\mathbb{l}(\check{c},\xi,\varphi,\psi)$ as the terms inside $e\mathbb{q}_\omega$ that contain $\check{c}$,\footnote{For the full expression, see \eqref{eq: full elambda}}  we have the following theorem.
\begin{theorem}
The collection $(\mathcal{F}_{SG},\varpi_{SG},Q,\mathcal{S})$ defines a BV structure, where $$\mathcal{S}=\mathcal{S}_2+\int_M \frac{1}{2}c^\dag\iota_\xi \mathbb{l}(\check{c},\xi,\varphi,\psi),$$   
and $\mathbb{l}(\check{c},\xi,\varphi,\psi)$ implicitly defined by 
    \begin{align*}
        \nonumber e\mathbb{l}(\check{c},\xi,\chi,\psi)\coloneqq&- \frac{1}{8\cdot 3!} \iota_\xi \check{c} \bar{\psi}\gamma^3 \vf\psi  -\frac{i}{4\cdot 3!} \bar{\psi}\gamma^3 \vf\left( \ugam\alpha(\check{c}\iota_\xi e \psi) \right)\\
        & -\frac{1}{2\cdot 3!}\bar{\psi}\gamma^3\chi\kappa\left[ <e, \bar{\chi}\left(  \frac{1}{2}\ugam\iota_\xi\check{c}\psi + \iota_\xi\ugam \check{c}\psi \right) > \right]\\
        &\nonumber- \frac{1}{16\cdot 3!}\bar{\psi}\gamma^3\chi \bar{\chi}\igam\igam\left(  \frac{1}{2}\ugam\iota_\xi\check{c}\psi + \iota_\xi\ugam \check{c}\psi  \right).
    \end{align*}
\end{theorem}
\begin{proof}
    The proof, which amounts to showing the classical master equation, is found in \ref{app: Q^2}.
\end{proof}

\section{Tools, lemmata and identities }
    This section provides some technical lemmata, useful throughout the paper.
    
    We start by defining the following spaces
        \begin{equation*}
            \Omega^{(k,l)}:=\Omega^k(M,\wedge^l \mathcal{V}) ,
        \end{equation*}
    and maps 
        \begin{eqnarray}
            W_{k}^{ (i,j)}: \Omega^{i,j}  \longrightarrow &\Omega^{i+k,j+k}  \label{map: W_e-bulk}\\
            X  \longmapsto  &  \frac{1}{k!}\underbrace{e \wedge \dots \wedge e}_{k-times}\wedge X  \nonumber
        \end{eqnarray}
        \begin{eqnarray}    
            \varrho^{i,j}: \Omega^{i,j}  \longrightarrow &\Omega^{i+1,j-1}  \label{map: [e,]-bulk}\\
            X \longmapsto & [e,X]\nonumber.
        \end{eqnarray}

            Such maps have been studied in previous papers (notably in \cite{CS2019,CCS2020} and \cite{Canepa:2024rib}. Here we provide a list of results and complement them with some new ones, which will be useful in the understanding of the boundary and bulk structure of supergravity.

        The following diagram \cite{Canepa:2024rib} indicates the properties of the maps $W_1^{(i,j)}$ and $W_1^{\partial,(i,j)} $, in particular a hooked arrow indicates injectivity while a two--headed arrow indicates surjectivity. In the bulk we have    
    \begin{equation}\label{diag: prop e bulk}
    \begin{tikzcd}
    	{\Omega^{(0,0)}} & {\Omega^{(0,1)}} & {\Omega^{(0,2)}} & {\Omega^{(0,3)}} & {\Omega^{(0,4)}} \\
    	{\Omega^{(1,0)}} & {\Omega^{(1,1)}} & {\Omega^{(1,2)}} & {\Omega^{(1,3)}} & {\Omega^{(1,4)}} \\
    	{\Omega^{(2,0)}} & {\Omega^{(2,1)}} & {\Omega^{(2,2)}} & {\Omega^{(2,3)}} & {\Omega^{(2,4)}} \\
    	{\Omega^{(3,0)}} & {\Omega^{(3,1)}} & {\Omega^{(3,2)}} & {\Omega^{(3,3)}} & {\Omega^{(3,4)}} \\
    	{\Omega^{(4,0)}} & {\Omega^{(4,1)}} & {\Omega^{(4,2)}} & {\Omega^{(4,3)}} & {\Omega^{(4,4)}}
    	\arrow[hook, from=2-1, to=3-2]
    	\arrow[hook, two heads, from=3-2, to=4-3]
    	\arrow[two heads, from=4-3, to=5-4]
    	\arrow[hook, from=1-1, to=2-2]
    	\arrow[hook, from=2-2, to=3-3]
    	\arrow[two heads, from=3-3, to=4-4]
    	\arrow[two heads, from=4-4, to=5-5]
    	\arrow[hook, from=1-2, to=2-3]
    	\arrow[hook, two heads, from=2-3, to=3-4]
    	\arrow[two heads, from=3-4, to=4-5]
    	\arrow[hook, from=1-3, to=2-4]
    	\arrow[two heads, from=2-4, to=3-5]
    	\arrow[hook, two heads, from=1-4, to=2-5]
    	\arrow[hook, from=3-1, to=4-2]
    	\arrow[two heads, from=4-2, to=5-3]
    	\arrow[hook, two heads, from=4-1, to=5-2]
    \end{tikzcd}        
    \end{equation}
   
    \begin{lemma}\label{lem: useful isos}
        The following maps are isomorphisms:
        \begin{enumerate}
            \item \label{lem: iso W2 (0,2)}$W_2^{(0,2)}\colon\Omega^{(0,2)}\rightarrow \Omega^{(2,4)} $,
            \item \label{lem: iso W2 (2,0)}$W_2^{(2,0) } \colon\Omega^{(2,0)}\rightarrow \Omega^{(4,2)} $,
            \item \label{lem: iso W2 (1,1)} $W_2^{(1,1) } \colon\Omega^{(2,0)}\rightarrow \Omega^{(3,3)} $,
            \item \label{lem: iso W4 (0,0)}$W_4^{(0,0) }\colon\Omega^{(0,0)}\rightarrow \Omega^{(4,4)} $,
            \item \label{lem: iso rho (0,1)}$\varrho^{(0,1)}\colon\Omega^{(0,1)}\rightarrow\Omega^{(1,0)} $.
            \item \label{lem: iso rho (3,4)}$\varrho^{(3,4)}\colon\Omega^{(3,4)}\rightarrow\Omega^{(4,3)} $.
        \end{enumerate}
    \end{lemma}
    \begin{proof}In all the above cases, the ranks\footnote{by this we mean the dimension of the fibers, e.g. dim$(\Omega^{(1,0)} ):=\dim( \Omega^{(1,0)}_x )=\dim(T^*_xM)=4$.} of the domain and target coincide, hence we only need to show that the appropriate $W_\bullet^{(\cdot,\cdot)}$ and $\varrho^{(\cdot,\cdot)}$ are injective.
    
        \begin{enumerate}
            \item Let $c\in\Omega^{(0,2)}$, then 
                \begin{align*}
                    \frac{e^2}{2}c\propto e^a_\mu e^b_\nu c^{df}dx^\mu dx^\nu v_a v_b v_d v_f = 0 \quad \Leftrightarrow \quad \epsilon_{abcd}e^a_{[\mu]} e^b_{\nu]} c^{df}=0.
                \end{align*}
                The above is a system of six independent equations, for $\mu,\nu=0,1,2,3$. One can easily prove that they uniquely fix the six components $c^{df}$ of $c$ to vanish.
            \item The proof is completely analogous to the above one, where the role of greek and latin indices is switched.
            \item Let $\theta \in \Omega^{(1,1)} $, then, similarly as before, one obtains
                \begin{equation*}
                    \frac{e^2}{2}\theta=0 \qquad \Leftrightarrow \qquad \epsilon_{abcd}e^a{[\mu}e^b_\nu \theta^c_{\rho]}=0,
                \end{equation*}
            which is a system of $4\times 4$ independent equations, whose unique  solution fixes the components $\theta^a_\mu$ of $\theta$ to vanish.   
            \item By explicit computation, $\frac{1}{4!}e^4=\text{Vol}_M$, hence the statement is immediately verified.
            \item Consider any $\alpha\in\Omega^{(0,1)}$, then 
                \begin{equation*}
                    [e,\alpha]=e^a_\mu \alpha^b \eta_{ab} dx^\mu = 0 \qquad \Leftrightarrow \qquad \alpha^b=0,
                \end{equation*}
            due to the invertibility of $e^a_\mu$.    
            \item Let $\Theta\in\Omega^{(3,4)}$, then by the above lemma there exist an $\alpha\in\Omega^{(0,1)} $ such that $\Theta=\frac{e^3}{3!}\alpha$, hence
                \begin{equation*}
                    [e,\Theta]=\frac{1}{3!}[e,e^3\alpha]= \frac{1}{2}e^2[e,e]\alpha + \frac{1}{3!}e^3[e,\alpha]=\frac{1}{3!}e^3[e,\alpha]=0 \quad \Leftrightarrow \quad \alpha=0,
                \end{equation*}
            having used $[e,e]=0$.
        \end{enumerate}
    \end{proof}

 \subsection{Results for gamma matrices and Majorana spinors in $D=4$}
    The following contains series of useful results, here without proofs, taken from \cite{FR25}.
    
    Before moving to the Majorana spinors, we take a look at some recurring identities regarding gamma matrices in $D=4$, having set $\gamma^5:=i\gamma^0\gamma^1\gamma^2\gamma^3$ and $\gamma_{a_1,\cdots,a_n}:= \gamma_{[a_1} \gamma_{a_2}\cdots \gamma_{a_n]}  $
    \begin{align}
        &\gamma^a\gamma^b\gamma^c\gamma^d\gamma_a=2\gamma^d\gamma^c\gamma^b;\\
        & \gamma^a\gamma^b\gamma^c=-\eta^{ab}\gamma^c-\eta^{bc}\gamma^a+\eta^{ac}\gamma^b + i \epsilon^{dabc}\gamma_d\gamma^5;\\
        &\gamma^5\gamma^{[c}\gamma^{d]}=-\frac{i}{2}\epsilon^{abcd}\gamma_{ab};\\
        &\label{id: gamma5gammac} \gamma^5\gamma^c=\frac{i}{6}\epsilon^{abcd}\gamma_{abc}.
    \end{align}
    Considering $\{v_a\}$ a basis for $V$, setting $\gamma:=\gamma^a v_a$ we have
        \begin{align*}
            [v_a,\cdot]\colon &\wedge^k V \longrightarrow  \wedge^{k-1}V\\
            &\alpha=\frac{1}{k!}\alpha^{a_1\cdots a_k}v_{a_1}\cdots v_{a_k} \longmapsto\frac{1}{(k-1)!}\eta_{a a_1} \alpha^{a_1\cdots a_k} v_{a_2}\cdots v_{a_k}.
        \end{align*}
    We obtain for all $c\in\wedge^2 V$
        \begin{align}
            & \label{id:v_a,gamma^N}[v_a,\gamma^N]=N[v_a,\gamma]\gamma^{N-1} + N(N-1)v_a\gamma^{N-2}.
        \end{align}
        
            \subsubsection{Majorana flip relations}
            Given any two Majorana spinors $\psi$ and $\chi$ of arbitrary parity, we have the following 
                \begin{align}
                    &  \label{flip:0}\bar{\chi}\psi= -(-1)^{|\chi||\psi|}\bar{\psi}\chi;\\
                    &  \label{flip:1}\bar{\chi}\gamma\psi=(-1)^{|\psi|+|\chi|+|\psi||\chi|}\bar{\psi}\gamma\chi; \\
                    & \label{flip:2} \bar{\chi}\gamma^2\psi=(-1)^{|\psi||\chi|}\bar{\psi}\gamma^2\chi  ;\\
                    &   \label{flip:3}\bar{\chi}\gamma^3\psi=-(-1)^{|\psi|+|\chi|+|\psi||\chi|}\bar{\psi}\gamma^3\chi.
                \end{align}
    
            In general, one finds 
                \begin{equation}
                    \label{flip:N}\bar{\chi}\gamma^N\psi=-t_N(-1)^{N(|\psi|+|\chi|)+|\psi||\chi|}\bar{\psi}\gamma^N\chi,
                \end{equation}
            where $t_N$ is defined from $(C\gamma^N)^t=-t_NC \gamma^N$ and is such that $t_{N+4}=t_N$. The first 4 parameters read
                \begin{equation*}
                    t_0=1,\qquad t_1=-1, \qquad t_2=-1, \qquad t_3=1.
                \end{equation*}
            Furthermore, another important identity derived from the ones above, is 
                \begin{equation}\label{id: action of omega}
                     \bar{\chi}\gamma[\alpha,\psi]=3\bar{\chi}\gamma\psi \alpha + \frac{1}{2}\bar{\chi}[\alpha,\gamma^3]_V\psi,
                \end{equation}
            which is true for all $\chi,\psi\in\mathbb{S}_M$  and $\alpha\in\wedge^2 V$.   
            \subsubsection{Fierz identities}
            Using the fact that the gamma matrices generate the whole Clifford algebra, which in the gamma representation is obtained as the algebra of matrices acting on $\mathbb{C}^4$, from the completeness relation one obtains the identity
                \begin{equation}\label{id: Fierz identiy}
                    (\gamma^a)^\cdot_{\alpha(\delta}(\gamma^a)^\cdot_{\rho\beta)}=0.
                \end{equation}
            Contracting with 4 Majorana spinors $\lambda_i$'s ($i=1, \cdots, 4$) of arbitary parity, we obtain the Fierz identities
                \begin{align}
                    & \label{Fierz:1}\bar{\lambda}_1 \gamma^3 \lambda_2 \bar{\lambda}_3 \gamma \lambda_4=(-1)^{|\lambda_2||\lambda_3|}\bar{\lambda}_1\gamma\lambda_3 \bar{\lambda}_2 \gamma^3 \lambda_4 + (-1)^{|\lambda_4|(|\lambda_2|+|\lambda_3|+1)+|\lambda_3|} \bar{\lambda}_1\gamma\lambda_4 \bar{\lambda}_2 \gamma^3 \lambda_3;\\
                    & \label{Fierz:2} \bar{\lambda}_1 \gamma^3 \lambda_2 \bar{\lambda}_3 \gamma \lambda_4= - (-1)^{|\lambda_2||\lambda_3|}\bar{\lambda}_1\gamma^3\lambda_3 \bar{\lambda}_2 \gamma \lambda_4 - (-1)^{|\lambda_4|(|\lambda_2|+|\lambda_3|+1)+|\lambda_3|} \bar{\lambda}_1\gamma^3\lambda_4 \bar{\lambda}_2 \gamma \lambda_3 .
                \end{align}

    \subsubsection{Lemmata}
    In the following we provide a series of lemmata holding for Majorana (and, when unspecified, Dirac) spinors. They appear verbatim, with proofs, in \cite{FR25}.
    \begin{lemma}\label{lem: injectivity egamma^3}
        The map 
        \begin{align*}
            \Theta^{(1,0)}\colon  \Omega^{(1,0)}(\Pi \mathbb{S}_D)& \longrightarrow \Omega^{(2,4)}(\Pi \mathbb{S}_D)\\
            \psi& \longmapsto \frac{1}{3!}e\gamma^3 \psi
        \end{align*}
        is injective.
    \end{lemma}

     \begin{lemma}\label{lem: iso (1,0) (3,4)}
        The map 
        \begin{align*}
            \Theta_\gamma^{(1,0)}\colon  \Omega^{(1,0)}(\Pi \mathbb{S}_D)& \longrightarrow \Omega^{(3,4)}(\Pi \mathbb{S}_D)\\
            \psi& \longmapsto \frac{1}{3!}e\gamma^3 \underline{\gamma}\psi
        \end{align*}
        is an isomorphism, where $\underline{\gamma}:= [e,\gamma]=\gamma_\mu dx^\mu$
    \end{lemma}

    \begin{remark}
        By the same reasoning (or just by taking the Dirac conjugate of the above expression), one finds that also the map
            \begin{equation*}
                 \psi \longmapsto \frac{1}{3!}e\ugam\gamma^3 \psi
            \end{equation*}
        is an isomorphism.    
    \end{remark}
    \begin{lemma}\label{lem: splitting (3,1)}
        For all $\theta \in \Omega^{(3,1)}(\Pi \mathbb{S}_M) $ there exist unique $\alpha\in \Omega^{(1,0)}(\Pi \mathbb{S}_M) $ and $\beta \in \Omega^{(3,1)}(\Pi \mathbb{S}_M) $ such that
            \begin{equation}
                \theta = i e \underline{\gamma} \alpha + \beta \qquad \mathrm{and}\qquad \gamma^3 \beta = 0.
            \end{equation}
    \end{lemma}
    
    \begin{lemma}\label{lem: splitting (2,1)}
        Let $\ugam\gamma^n_{(i,j)}$ be the map
            \begin{equation*}
                \ugam\gamma^n_{(i,j)}\colon\Omega^{(i,j)}\rightarrow \Omega^{(i+1,j+n)}\colon\beta\mapsto \ugam\gamma^n \beta,
            \end{equation*}
        then, for all $\theta \in \Omega^{(2,1)}(\Pi \mathbb{S}_M)$ there exist unique $\alpha\in\Omega^{(1,0)}(\Pi \mathbb{S}_M)$ and $\beta \in \ker{\ugam\gamma^3_{(2,1)}}$ such that 
            \begin{equation*}
                \theta= e \alpha + \beta. 
            \end{equation*}
    \end{lemma}
 
    \begin{definition}
        Thanks to the previous lemmata, we can define maps, 
        \begin{align*}
            &\alpha\colon\Omega^{(3,1)}(\Pi \mathbb{S}_M)\rightarrow \Omega^{(1,0)}(\Pi \mathbb{S}_M) &\beta\colon\Omega^{(3,1)}(\Pi \mathbb{S}_M)\rightarrow\ker{(\gamma^3_{(3,1)})}\\
            &\kappa\colon\Omega^{(2,1)}(\Pi \mathbb{S}_M)\rightarrow\Omega^{(1,0)}(\Pi \mathbb{S}_M) & \varkappa\colon \Omega^{(2,1)}(\Pi \mathbb{S}_M)\rightarrow\ker{(\ugam\gamma^3_{(2,1)})}
        \end{align*}
        such that for all $\theta\in\Omega^{(3,1)} $ and $\omega\in\Omega^{(2,1)}$, we have
            \begin{align*}
               & \theta= i e \ugam\alpha(\theta) + \beta(\theta), \qquad\omega = e \kappa(\omega) + \varkappa(\omega).
            \end{align*}
    \end{definition}
    
    \begin{lemma}\label{lem: Fierz}
        For all $\lambda,\psi,\chi\in \mathbb{S}_M$  such that $|\chi|=0$ and $|\psi|=1$, the following identities hold
            \begin{equation*}
                \bar{\lambda}\gamma^3 \chi \bar{\chi}\gamma\psi =0, \qquad  \bar{\chi}\gamma\chi \bar{\lambda}\gamma^3 \psi=0 \qquad \mathrm{and} \qquad \bar{\lambda}\gamma\chi \bar{\chi}\gamma^3\psi=0.
            \end{equation*}
    \end{lemma}

\section{Lengthy computations of section \ref{sec: BV formalism} }

\subsection{Computation of $Q_0^2$}\label{app: Q_0^2}
To make computations easier, we can split $Q_0=Q_{PC}+\delta_\chi$, where $Q_{PC}$ is the cohomological vector field associated to the BV theory of pure Palatini--Cartan gravity. We then start by computing $\delta_\chi^2$ on the fields and ghosts. We see
\begin{align*}
    \delta_\chi^2 e &=\delta_\chi(\bar{\chi}\gamma \psi)=-\frac{1}{2}\vf \bar{\psi}\gamma \psi + \bar{\chi}\gamma d_\omega \psi = -\frac{1}{2}d_\omega(\bar{\chi}\gamma\chi) + \frac{1}{2}\vf\left( -\frac{1}{2}\bar{\psi}\gamma\psi \right)\\
    &=-\frac{1}{2}\mathrm{L}^\omega_\varphi e + \frac{1}{2}\vf\left( d_\omega e  -\frac{1}{2}\bar{\psi}\gamma\psi  \right),
\end{align*}
having used $\vf e =\bar{\chi}\gamma \chi$. To compute $Q_0^2 e,$ we simply use $Q_0^2=[Q_PC,\delta_\chi]+\delta_\chi^2$, obtaining
\begin{align*}
    Q_0^2 e &= Q_{PC}(\bar{\chi}\gamma\psi) + \delta_\chi(\mathrm{L}^\omega_\xi e -[c,e] ) -\frac{1}{2}\mathrm{L}^\omega_\varphi e + \frac{1}{2}\vf\left(\mathrm{EoM}_\omega  \right)\\
    &= \mathrm{L}^\omega_\xi\bar{\chi}\gamma\psi - [c,\bar{\chi}]\gamma\psi -\bar{\chi}\gamma \mathrm{L}^\omega_\xi \psi + \bar{\chi}\gamma[c,\psi] + \frac{1}{2}\mathrm{L}^\omega_\varphi e +[\iota_\xi \omega_\chi,e] \\
    &\quad -\mathrm{L}^\omega_\xi (\bar{\chi} \gamma\psi) -[\iota_\xi\delta_\chi\omega,\chi] + [c,\bar{\chi}\gamma\psi]  -\frac{1}{2}\mathrm{L}^\omega_\varphi e + \frac{1}{2}\vf\left(\mathrm{EoM}_\omega  \right)\\
    &=\frac{1}{2}\vf\left(\mathrm{EoM}_\omega  \right).
\end{align*}
We now move to the computation of $\delta_\chi^2 \psi$,
\begin{align*}
    \delta_\chi^2 \psi &=\delta_\chi(-d_\omega \chi )=-[\delta_\chi \omega,\chi] -\frac{1}{2}d_\omega \vf \psi=-[\delta_\chi \omega,\chi] - \frac{1}{2}\mathrm{L}_\varphi^\omega \psi + \frac{1}{2}\vf d_\omega\psi.
\end{align*}
In order to continue, we need to compute the explicit form of $\delta_\chi\omega$. In order to do so, we rewrite $d_\omega \psi$ in the veilbein basis, i.e. define
    \begin{equation*}
        d_\omega \psi:=\frac{1}{2}\rho_{ab}e^a e^b=\frac{1}{4}[v_a,[v_b,e^2]]\rho^{ab},
    \end{equation*}
hence obtaining
    \begin{align*}
        e\delta_\chi\omega&=-\frac{1}{4\cdot 3!}\bar{\chi}[v_a,[v_b,e^2]]\gamma^3\rho^{ab}=-\frac{1}{4\cdot 3!}\bar{\chi}\left([v_a,e^2 [v_b,\gamma^3]] -[v_b,e^2][v_a,\gamma^3] \right)\\
        &=e\bar{\chi}\left( \frac{1}{2}\gamma_a \gamma^2 e_b -\gamma v_a e_b + \frac{1}{4}e\gamma_b\gamma_a \gamma +\frac{1}{2} e v_a \gamma^b \right)\rho^{ab}.
    \end{align*}
Defining $\hat{\gamma}:=\gamma^a e_a^\mu \partial_\mu\in \mathfrak{X}(M)$, and the map\footnote{Notice that, with this definition, $[e,<\hat{e},d_\omega\psi>]=2d_\omega\psi.$}
    \begin{align*}
        <\bar{e},\cdot>\colon \Omega^{(i,j)}&\rightarrow\Omega^{(i-1,j+1)}\\
        \alpha & \mapsto v_a \eta^{ad}e_d^\mu \iota_{\partial_\mu}\alpha,
    \end{align*} we have
    \begin{equation}
        \delta_\chi\omega=\frac{1}{2}\bar{\chi}\iota_{\hat{\gamma}}(\gamma^2 d_\omega\psi) - \bar{\chi}\gamma <\bar{e},d_\omega\psi>+\frac{1}{4}e\bar{\chi}\iota_{\hat{\gamma}}\iota_{\hat{\gamma}}(\gamma d_\omega\psi) - \frac{1}{2}e \bar{\chi}\iota_{\hat{\gamma}}<\bar{e},d_\omega\psi>.
    \end{equation}
Now, for computational reasons, it is easier to compute $\frac{1}{3!}e\ugam\gamma^3  [\delta_\chi \omega,\chi]$, since it provides an isomorphism from lemma \ref{lem: iso (1,0) (3,4)}. Furthermore, we can without loss of generality contract the expression with respect to a generic Majorana spinor $\lambda$. Using \eqref{id: action of omega}, we find
\begin{align*}
    \frac{1}{3!}e\bar{\lambda}\ugam\gamma^3  [\delta_\chi \omega,\chi]&= \frac{1}{2}\bar{\lambda}\ugam \gamma \chi e\delta_\chi \omega + \frac{1}{2\cdot 3!} e \bar{\lambda}\ugam[\delta_\chi \omega,\gamma^3]\chi\\
    &= - \frac{1}{2\cdot 3! } \bar{\lambda}\ugam\gamma\chi \bar{\chi}\gamma^3 d_\omega\psi + \frac{1}{2\cdot 3!}\bar{\lambda}\ugam\gamma^3 \chi [e,\delta_\chi \omega]\\
    &\overset{\eqref{lem: Fierz}}{=}\frac{1}{2\cdot 3!}\bar{\lambda}\ugam\gamma^3 \chi [e,\delta_\chi \omega].
\end{align*}
Now, a rather long but straightforward computation gives
    \begin{align*}
        [e,\delta_\chi\omega]&= 5\bar{\chi}\gamma d_\omega\psi + \bar{\chi }\gamma\iota_{\hat{\gamma}}(\ugam d_\omega\psi) + \bar{\chi}<\bar{e},\ugam d_\omega\psi>  + \frac{1}{4} e\bar{\chi}\iota_{\hat{\gamma}}\iota_{\hat{\gamma}}(\ugam d_\omega\psi).
    \end{align*}
We notice that in $\bar{\lambda}\ugam\gamma^3 \chi [e,\delta_\chi \omega]$ all the terms containing $\bar{\chi}\gamma(\cdot)$ vanish because of lemma \ref{lem: iso (1,0) (3,4)}. Hence, eliminating the arbitary spinor $\lambda$, we are left with
\begin{align*}
    \frac{1}{3!}e\ugam\gamma^3  [\delta_\chi \omega,\chi]&= \frac{1}{2\cdot 3!}\ugam\gamma^3 \chi \bar{\chi}<\bar{e},\ugam d_\omega\psi> +\frac{1}{8\cdot 3!}e \ugam\gamma^3 \chi\bar{\chi}\iota_{\hat{\gamma}}\iota_{\hat{\gamma}}(\ugam d_\omega\psi)\\
    &\overset{\ref{lem: splitting (2,1)}}{=} \frac{1}{3!}e \ugam \gamma^3 \chi \left( \bar{\chi} \kappa(<\bar{e},\ugam d_\omega \psi>) + \frac{1}{8} \bar{\chi} \iota_{\hat{\gamma}}\iota_{\hat{\gamma}}(\ugam d_\omega\psi) \right),
\end{align*}
hence showing 
    \begin{equation*}
        [\delta_\chi\omega,\chi]=  \left(\bar{\chi} \kappa(<\bar{e},\ugam d_\omega \psi>) + \frac{1}{8} \bar{\chi} \iota_{\hat{\gamma}}\iota_{\hat{\gamma}}(\ugam d_\omega\psi) \right)\chi,
    \end{equation*}
and
    \begin{equation*}
        \delta_\chi^2 \psi = - \frac{1}{2}\mathrm{L}_\varphi^\omega \psi + \frac{1}{2}\vf d_\omega\psi -  \left(\bar{\chi} \kappa(<\bar{e},\ugam d_\omega \psi>) + \frac{1}{8} \bar{\chi} \iota_{\hat{\gamma}}\iota_{\hat{\gamma}}(\ugam d_\omega\psi) \right)\chi.
    \end{equation*}
We therefore obtain
    \begin{align*}
        Q_0^2\psi&=Q_0(\mathrm{L}^\omega_\xi\psi -[c,\psi] - d_\omega\chi)\\
        &=\frac{1}{2}\mathrm{L}^\omega_{[\xi,\xi]}\psi + \frac{1}{2}\mathrm{L}^\omega_\varphi\psi + [\iota_\xi\iota_\xi F_\omega,\psi] -[\mathrm{L}^\omega_\xi c,\psi] + [\iota_\xi\delta_\chi\omega,\psi]\\
        &\quad - \mathrm{L}^\omega_\xi\mathrm{L}^\omega_\xi\psi +\mathrm{L}^\omega_\xi[c,\psi] + \mathrm{L}^\omega_\xi d_\omega\chi +\frac{1}{2} [[c,c],\psi] - \frac{1}{2}[\iota_\xi\iota_\xi F_\omega,\psi]  \\
        &\quad  - [\iota_\xi\delta_\chi\omega,\psi] + [c, \mathrm{L}^\omega_\xi\psi] -[c,[c,\psi]] - [c,d_\omega\chi] +\delta_\chi^2\psi \\
        &\quad  -[\iota_\xi F_\omega,\chi] + [ - \mathrm{L}^\omega_\xi c,\chi] + d_\omega\left(  - \mathrm{L}^\omega_\xi\chi -[c,\chi]  \right) \\
        &=\frac{1}{2}\vf d_\omega\psi -    \left(\bar{\chi} \kappa(<\bar{e},\ugam d_\omega \psi>) + \frac{1}{8} \bar{\chi} \iota_{\hat{\gamma}}\iota_{\hat{\gamma}}(\ugam d_\omega\psi) \right)\chi,
    \end{align*}
having noticed the following
\begin{itemize}
    \item $\frac{1}{2}\mathrm{L}^\omega_{[\xi,\xi]}\psi + [\iota_\xi\iota_\xi F_\omega,\psi]  - \mathrm{L}^\omega_\xi\mathrm{L}^\omega_\xi\psi  - \frac{1}{2}[\iota_\xi\iota_\xi F_\omega,\psi] =0 $;
    \item $ \mathrm{L}^\omega_\xi d_\omega \chi + d_\omega\mathrm{L}^\omega_\xi\chi - [\iota_\xi F_\omega,\chi]=0 $, since $[\mathrm{L}^\omega_\xi,d_\omega]=[\iota_\xi F_\omega,\cdot]$ on any field;
    \item $ \frac{1}{2} [[c,c],\psi] - [c,[c,\psi]]=0 $ using graded Jacobi identity.
\end{itemize}
Regarding $\delta_\chi^2\omega$, we first start by noticing that, as stated before, computing $e\delta_\chi^2\omega$ defines it univocally. Hence we obtain
    \begin{align*}
        e\delta_\chi^2\omega&=\delta_\chi(e\delta_\chi\omega) - \bar{\chi}\gamma\psi \delta_\chi\omega= \delta_\chi\left( -\frac{1}{3!}\bar{\chi}\gamma^3 d_\omega\psi \right)- \bar{\chi}\gamma\psi \delta_\chi\omega\\
        &= \frac{1}{2\cdot 3!}\vf(\bar{\psi}) \gamma^3 d_\omega\psi + \frac{1}{3!}\bar{\chi}\gamma^3 [\delta_\chi \omega,\psi] + \frac{1}{3!}\bar{\chi}\gamma^3 [F_\omega,\chi]  - \bar{\chi}\gamma\psi \delta_\chi\omega\\
        &= \frac{1}{2}\vf\left( \frac{1}{3!}\bar{\psi}\gamma^3 d_\omega \psi \right) - \frac{1}{2\cdot 3!}\bar{\psi}\vf(\gamma^3 d_\omega\psi)  + \frac{1}{3!}\bar{\chi}\gamma^3 [\delta_\chi \omega,\psi]\\
        &\quad +\frac{1}{2}F_\omega \bar{\chi}\gamma\chi- \bar{\chi}\gamma\psi \delta_\chi\omega\\
        &=-\frac{1}{2}e\vf F_\omega +\frac{1}{2}\vf\left(  eF_\omega+ \frac{1}{3!}\bar{\psi}\gamma^3 d_\omega \psi \right) - \frac{1}{2\cdot 3!}\bar{\psi}\vf(\gamma^3 d_\omega\psi) \\
        &\quad + \frac{1}{3!}\bar{\chi}\gamma^3 [\delta_\chi \omega,\psi] - \bar{\chi}\gamma\psi \delta_\chi\omega.
    \end{align*}
Now we can use \eqref{id: action of omega} to see
    \begin{equation*}
       \bar{\chi}\gamma\psi \delta_\chi\omega=\frac{1}{3!}\bar{\chi}\gamma^3[\delta_\chi\omega,\psi]-\frac{1}{3!}[\delta_\chi\omega,\bar{\chi}]\gamma^3\psi,
    \end{equation*}
hence 
    \begin{align*}
        e\delta_\chi^2\omega=&-\frac{1}{2}e\vf F_\omega +\frac{1}{2}\vf\left(  eF_\omega+ \frac{1}{3!}\bar{\psi}\gamma^3 d_\omega \psi \right) - \frac{1}{2\cdot 3!}\bar{\psi}\vf(\gamma^3 d_\omega\psi)
        -\frac{1}{3!}\bar{\psi}\gamma^3[\delta_\chi\omega,\chi]\\
        =&-\frac{1}{2}e\vf F_\omega +\frac{1}{2}\vf\left(  eF_\omega+ \frac{1}{3!}\bar{\psi}\gamma^3 d_\omega \psi \right) - \frac{1}{2\cdot 3!}\bar{\psi}\vf(\gamma^3 d_\omega\psi)\\
        &-\frac{1}{3!}\bar{\psi}\gamma^3\chi\left(\bar{\chi} \kappa(<\bar{e},\ugam d_\omega \psi>) + \frac{1}{8} \bar{\chi} \iota_{\hat{\gamma}}\iota_{\hat{\gamma}}(\ugam d_\omega\psi) \right).
    \end{align*}
We  see
    \begin{align*}
        eQ_0(\delta_\chi\omega)&= -\frac{1}{3!}Q_0(\bar{\chi}\gamma^3d_\omega\psi) - (\mathrm{L}^\omega_\xi e - [c,e])\delta_\chi \omega + e \delta_\chi^2\omega\\
        &= -\frac{1}{3!}(\mathrm{L}^\omega_\xi \bar{\chi} -[c,\bar{\chi}] )\gamma^3d_\omega\psi +\frac{1}{3!}\bar{\chi}\gamma^3 [\iota_\xi F_\omega -d_\omega c,\psi] \\
        &\quad - \frac{1}{3!}\bar{\chi}\gamma^3d_\omega(\mathrm{L}^\omega_\xi \psi - [c,\psi]) - (\mathrm{L}^\omega_\xi e - [c,e])\delta_\chi \omega + e \delta_\chi^2\omega\\
        &= -\frac{1}{3!}\mathrm{L}^\omega_\xi(\bar{\chi}\gamma^3d_\omega\psi) +\frac{1}{3!}[c,\bar{\chi}\gamma^3d_\omega\psi]  - (\mathrm{L}^\omega_\xi e - [c,e])\delta_\chi \omega + e \delta_\chi^2\omega \\
        &=  e (\mathrm{L}^\omega_\xi \delta_\chi\omega -[c,\delta_\chi\omega]) + e \delta_\chi^2\omega,
    \end{align*}
hence obtaining
    \begin{align*}
        eQ_0^2\omega&= eQ_0(\iota_\xi F_\omega -d_\omega c + \delta_\chi\omega )\\
        &= \frac{1}{2}e\vf F_\omega -e\iota_\xi d_\omega\delta_\chi\omega - e[\delta_\chi\omega,c] +e d_\omega \iota_\xi \delta_\chi\omega +e\mathrm{L}^\omega_\xi \delta_\chi\omega \\
        &\quad -e[c,\delta_\chi\omega] + e \delta^2_\chi\omega\\
        &=  \frac{1}{2}\vf\left( \mathrm{EoM}_e \right) - \frac{1}{2\cdot 3!}\bar{\psi}\vf(\gamma^3 d_\omega\psi)\\
        &\quad-\frac{1}{3!}\bar{\psi}\gamma^3\chi\left(\bar{\chi} \kappa(<\bar{e},\ugam d_\omega \psi>)+ \frac{1}{8} \bar{\chi} \iota_{\hat{\gamma}}\iota_{\hat{\gamma}}(\ugam d_\omega\psi) \right)
    \end{align*}
For $c$, $\chi$ and $\xi$, we can do the computations of $Q_0^2$ right away, obtaining
    \begin{align*}
        Q_0^2 c &= Q_{PC}(\iota_\xi \delta_\chi \omega) + \delta_\chi\left( \frac{1}{2}\iota_\xi\iota_\xi F_\omega - \frac{1}{2}[c,c] + \iota_\xi \delta_\chi \omega \right)\\
        &= \frac{1}{2}\iota_{[\xi,\xi]}\delta_\chi\omega + \iota_\xi\mathrm{L}^\omega_\xi \delta_\chi\omega -\iota_\xi[c,\delta_\chi\omega] + \frac{1}{2}\iota_\xi\vf F_\omega - \frac{1}{2}\iota_\xi\iota_\xi d_\omega \delta_\chi\omega \\
        &\quad - [\iota_\xi \delta_\chi\omega,c]   + \frac{1}{2}\iota_{\varphi}\delta_\chi\omega +\iota_\xi \delta_\chi^2\omega\\
        &=\frac{1}{2}\iota_{\varphi}\delta_\chi\omega +\iota_\xi Q_0^2\omega,
    \end{align*}
having used the fact that $\frac{1}{2}\iota_{[\xi,\xi]}\delta_\chi\omega + \iota_\xi\mathrm{L}^\omega_\xi \delta_\chi\omega - \frac{1}{2}\iota_\xi\iota_\xi d_\omega \delta_\chi\omega=0 $.

For $Q_0^2\xi$, we see    
    \begin{align*}
        Q_0^2(\xi)&=\frac{1}{2}Q_0([\xi,\xi]+\varphi) = \frac{1}{2}[\xi,\varphi] + \frac{1}{2}Q_0(\bar{\chi}\gamma^a\chi e_a^\mu)\partial_\mu\\
        &= \frac{1}{2}[\xi,\varphi] + \mathrm{L}_\xi^\omega(\bar{\chi})\gamma^\mu\chi \partial_\mu - [c,\bar{\chi}]\gamma^\mu\chi \partial_\mu -\frac{1}{2}\vf\bar{\psi}\gamma^\mu\chi \partial_\mu +\frac{1}{2}\bar{\chi}\gamma^a \chi Q_0(e_a^\mu)\partial_\mu.
    \end{align*}
Now, since $e_a^\mu e_\mu^b=\delta^b_a$, we have $Q_0(e^\mu_a)=-e^\nu_a e_b^\mu Q_0(e_\nu^b)$, obtaining
    \begin{align*}
        Q_0^2(\xi)
        &= \frac{1}{2}[\varphi,\xi] + \mathrm{L}_\xi^\omega(\bar{\chi})\gamma^\mu\chi \partial_\mu - [c,\bar{\chi}]\gamma^\mu\chi \partial_\mu -\frac{1}{2}\vf\bar{\psi}\gamma^\mu\chi \partial_\mu \\
        &\quad-\frac{1}{2}e^\mu_b  \bar{\chi}\gamma^\nu \chi\left( (\mathrm{L}_\xi^\omega e)^b_\nu -[c,e_\nu]^b + \bar{\chi} \gamma^b \psi_\nu \right)\partial_\mu\\
        &=-\frac{1}{2}[\xi,\varphi]+ \mathrm{L}_\xi^\omega(\bar{\chi})\gamma^\mu\chi \partial_\mu- [c,\bar{\chi}]\gamma^\mu\chi \partial_\mu -\frac{1}{2}\vf\bar{\psi}\gamma^\mu\chi \partial_\mu \\
        &\quad+\frac{1}{2}\bar{\chi}[c,\gamma]_V^\mu \chi\partial_\mu -\frac{1}{2}\bar{\chi}\gamma^\mu \vf\psi \partial_\mu +\frac{1}{2}\bar{\chi}\mathrm{L}_\xi^\omega(\gamma^\mu\partial_\mu)\chi\\
        &=-\frac{1}{2} \mathrm{L}_\xi^\omega(\phi)+\frac{1}{2}\mathrm{L}_\xi^\omega(\bar{\chi}\gamma^\mu\chi\partial_\mu) =0.
    \end{align*}
Notice also that this tells us that $Q_0\varphi=\mathrm{L}_\xi^\omega(\varphi)=[\xi,\varphi]$. Lastly,
    \begin{align*}
        Q_0^2\chi&= Q_0\left(\mathrm{L}_\xi^\omega\chi - [c,\chi] -\frac{1}{2}\vf\psi \right)\\
        &= \frac{1}{2}\mathrm{L}_{[\xi,\xi]}^\omega\chi + \frac{1}{2}\mathrm{L}_\varphi^\omega\chi +[\iota_\xi\iota_\xi F_\omega,\chi] -[\mathrm{L}_\xi^\omega c,\chi] + [\iota_\xi \delta_\chi \omega, \chi] -\mathrm{L}_\xi^\omega\mathrm{L}_\xi^\omega\chi\\
        &\quad +\mathrm{L}_\xi^\omega[c,\chi] +\frac{1}{2}\mathrm{L}_\xi^\omega\vf\psi - \frac{1}{2}[\iota_\xi\iota_\xi F_\omega,\chi] + \frac{1}{2}[[c,c],\chi] - [\iota_\xi \delta_\chi \omega,\chi] \\
        &\quad + [c,\mathrm{L}_\xi^\omega \chi] - [c,[c,\chi]] - \frac{1}{2}[c,\vf\psi] -\frac{1}{2}\iota_{[\xi,\varphi]}\psi +\frac{1}{2}\vf\left( \mathrm{L}_\xi^\omega\psi -[c,\psi] - d_\omega\chi \right)\\
        &=0.
    \end{align*}

Lastly, for the sake of completeness, we also provide the expression of $Q_0$ for the anti--fields. It is obtained by computing $\delta_{\mathrm{fields}}\mathcal{S}_1$. In particular, as we saw in section \ref{sec: BV formalism}, $Q_0\Phi^\dag$ will be proportional to the equations of motion for the respective fields, and for $c^\dag$ and $\omega^\dag$ we will have $Q_0=Q_{PC}+ \delta_\chi$. We compute $\delta_\chi$ as the Hamiltonian vector field of $s_1$, namely such that $\iota_{\delta_\chi}\varpi_{BV}=\delta s_1$, using 
    \begin{align*}
        \iota_{\mathbb{X}}\varpi_{BV}&=\int_M \mathbb{X}_e \delta e^\dag + \delta e ( \mathbb{X}_{e^\dag} + \mathbb{X}_\omega \check{\omega} + e\mathbb{X}_c \check{c} ) + e\mathbb{X}_\omega \delta\check{\omega} + \delta \omega(e \mathbb{X}_\omega + 
        \check{\omega}\mathbb{X}_e)\\
        &\qquad + i\mathbb{X}_{\bar{\psi}}\delta\psi_\dag + i \delta{\bar{\psi}}\mathbb{X}_{\psi_\dag} + \frac{e^2}{2}\mathbb{X}_c \delta \check{c} + \delta c \left( \frac{e^2}{2}\mathbb{X}_{\check{c}} + e \check{c}\mathbb{X}_e  \right)\\
        &\qquad + \iota_{\mathbb{X}_\xi}\delta \xi^\dag + \iota_{\delta\xi}\mathbb{X}_{\xi^\dag}  + i\mathbb{X}_{\bar{\chi}}\delta\chi_\dag + i \delta{\bar{\chi}}\mathbb{X}_{\chi_\dag},
    \end{align*}
Furthermore, one can also split $s_1=s_1^{PC}+s_1^\chi$, where $s_1^{PC}$ is the part coming from the free gravity BV theory. We are then left with
finding
    \begin{align*}
        &Q_0 e^\dag=e F_\omega +\frac{1}{3!}\bar{\psi}\gamma^3 d_\omega\psi +\mathrm{L}_\xi^\omega e^\dag -[c,e^\dag] -\frac{i}{2\cdot 3!}\vf(e^3\bar{\psi})\chi^0_\dag +\frac{1}{2}\vf[v_c,e^\mu_b\eta^{bc}\xi^\dag_\mu] \\
        &\hspace{11mm} -\frac{1}{2\cdot 3!}\bar{\chi}\gamma^3 d_\omega\psi +\frac{1}{3!}\iota_\xi(\bar{\chi}\gamma^3 d_\omega\psi \check{c} ) -\check{\omega}\delta_\chi\omega - e \check{c}\iota_\xi\delta_\chi\omega  \\
        &e Q_0\check{\omega}=e\left(d_\omega e -\frac{1}{2}\bar{\psi}\gamma\psi \right)-\iota_\xi[e^\dag,e] - d_\omega(\iota_\xi \omega^\dag) - e[c,\check{\omega}] +\frac{1}{2}d_\omega\iota_\xi\iota_\xi c^\dag -\check{\omega}\mathrm{L}^\omega_\xi e \\
        &\hspace{11mm} -\frac{1}{2}\bar{\chi}\gamma\psi \check{\omega}-\frac{1}{2\cdot3!}\bar{\chi}[\check{\omega},\gamma^3]_V\psi -\frac{1}{2}\bar{\chi}\gamma\psi\left( \check{c}\iota_\xi e +\frac{1}{2}e \iota_\xi \check{c} \right) +\frac{1}{2\cdot 3!}\bar{\chi}[ \check{c}\iota_\xi e +\frac{1}{2}e \iota_\xi \check{c} ,\gamma^3 ]\psi \\
        &\hspace{11mm} +\frac{i}{2}\iota_\xi\left( e\bar{\psi}^0_\dag \ugam\gamma\psi - \frac{1}{3!}{\psi}^0_\dag \ugam[e,\gamma^3]\psi \right) +\frac{i}{2}e\bar{\psi}^0_\dag\ugam\gamma\chi -\frac{i}{2\cdot 3!}\bar{\psi}^0_\dag \ugam[e,\gamma^3]\chi +\frac{i}{8}\iota_\xi\left(e^2 \bar{\chi}^0_\dag \ugam^2\chi \right)  \\
        &Q_0 \psi_\dag= -\frac{i}{3}\left( e\gamma^3 d_\omega\psi -\frac{1}{2}d_\omega e \gamma^3 \psi \right) -\frac{i}{3!} \gamma^3 d_\omega(\check{\omega}\chi) + \mathrm{L}^\omega_\xi\psi_\dag -[c,\psi_\dag] +i\gamma\chi e^\dag \\
        &\hspace{11mm} -\frac{i}{3!}d_\omega\left( \check{c}\iota_\xi e \gamma^3\chi + \frac{1}{2}e \iota_\xi \check{c}\gamma^3 \chi \right) -\frac{1}{2}\vf\chi_\dag  \\
        & \frac{e^2}{2}Q_0 \check{c}= -d_\omega\omega^\dag -[e,e^\dag]  +\frac{i}{8}e^2\bar{\chi}^0_\dag\ugam^2\chi + \frac{i}{2} e \bar{\psi}^0_\dag\ugam\gamma\psi -\frac{i}{2\cdot 3!} \bar{\psi}^0_\dag\ugam[e,\gamma^3]\psi -\frac{1}{2}\check{c}\mathrm{L}_\xi^\omega e^2 - \check{c}e\bar{\chi}\gamma\psi,\\
        &Q_0{\xi^\dag_\bullet}= - e^\dag_\bullet d_\omega e - d_\omega e e^\dag_\bullet - \omega^\dag_\bullet F_\omega -(\iota_\xi c^\dag)_\bullet F_\omega + \iota_{[\bullet,\xi]}\xi^\dag - i d\omega\bar{\psi}(\psi_\dag)_\bullet \\
        &\hspace{10mm}+ i\bar{\psi}_\bullet d_\omega\psi_\dag - i (d_\omega \bar{\chi})_\bullet \chi_\dag+ \frac{1}{ 3!}\check{c}e_\bullet\bar{\chi}\gamma^3d_\omega\psi + \frac{1}{2\cdot 3!}e\check{c}_\bullet\bar{\chi}\gamma^3d_\omega\psi\\
        &Q_0 \chi_\dag= \frac{i}{3!}\gamma^3 d_\omega\psi + i\gamma\psi e^\dag - d_\omega\psi_\dag -\igam\xi^\dag\chi,
    \end{align*}
having used \ref{lem: useful isos}.\ref{lem: iso W4 (0,0)} to redefine 
\begin{equation}
    \chi_\dag=\frac{e^4}{4!}\chi^0_\dag.
\end{equation}
    
\subsection{Computation of $Q^2$}\label{app: Q^2}    
First of all, we recall the quadratic part of the action $s_2$ from \eqref{eq: quadratic BV action}
    \begin{align*}
        \nonumber s_2&=\int_M \frac{1}{2}\left(\check{\omega}-\frac{1}{2}e\iota_\xi \check{c} - \check{c}\iota_\xi e\right)\vf e^\dag +\frac{1}{4}\left( \frac{1}{2}\bar{\psi}^0_\dag\ugam + \alpha(\check{\omega}\bar{\psi})\ugam -\frac{i}{2}\iota_\xi\check{c}\bar{\psi}-\alpha(\check{c}\iota_\xi e \bar{\psi})\ugam -\frac{i}{2}\check{c}\bar{\chi}\right)\vf\psi_\dag\\
        \nonumber&\qquad + \frac{i}{4\cdot 3!}\left(\frac{1}{2}\alpha(\check{\omega}\bar{\psi})\ugam -\frac{i}{2}\iota_\xi\check{c}\bar{\psi}-\alpha(\check{c}\iota_\xi e \bar{\psi})\ugam -\frac{i}{2}\check{c}\bar{\chi}\right)\gamma^3\vf(\check{\omega}\psi)\\
        &\qquad -\frac{i}{2\cdot 3!}\left( \frac{1}{2}\bar{\psi}^0_\dag\ugam+\frac{1}{2}\alpha(\check{\omega}\bar{\psi})\ugam -\frac{i}{2}\iota_\xi\check{c}\bar{\psi}-\alpha(\check{c}\iota_\xi e \bar{\psi})\ugam \right)\gamma^3\chi <e,\bar{\chi}[\check{\omega},\gamma]\psi>\\
        \nonumber&\qquad +\frac{1}{2\cdot 3!}\left( \frac{1}{4}\bar{\psi}^0_\dag\ugam-\frac{i}{2}\iota_\xi\check{c}\bar{\psi}-\alpha(\check{c}\iota_\xi e \bar{\psi})\ugam \right)\gamma^3\chi <e,\bar{\chi}\ugam^2\psi^0_\dag>\\
        \nonumber&\qquad -\frac{1}{32}\left(i\bar{\psi}_\dag \chi +\frac{1}{3!}(\check{\omega} - e \iota_\xi \check{c} - 2 \check{c}\iota_\xi e  )\bar{\psi}\gamma^3 \chi \right)\bar{\chi}\igam\igam([\check{\omega},\gamma]\psi)\\
        \nonumber&\qquad - \frac{i}{32}\left(i\bar{\psi}_\dag\chi + \frac{1}{3!}(e\iota_\xi \check{c} + 2\check{c}\iota_\xi e )\bar{\psi}\gamma^3\chi  \right)\bar{\chi}\igam\igam(\ugam^2\psi^0_\dag).
    \end{align*}
    
The variation of $s_2$ is long and tedious, hence we do not provide the details. However, carefully carrying out the computation yields the following Hamiltonian vector fields 
    \begin{align*}
         &\mathbb{q}_e =  \frac{1}{2}\iota_\varphi\check{\omega} -\frac{1}{2}\iota_\varphi\check{c}\iota_\xi e -\frac{1}{4}\iota_\varphi(e \iota_\xi \check{c} ) \\
         &e \mathbb{q}_\omega =   \frac{1}{2} \vf e^\dag +\frac{i}{4\cdot 3!} \vf(\bar{\psi^0_\dag}\ugam)\gamma^3 \psi +\frac{i}{4\cdot 3! }\bar{\psi}\gamma^3\vf\left(\ugam\alpha(\check{\omega}\psi )\right) - \frac{1}{8\cdot 3!} \vf \check{c} \bar{\chi}\gamma^3 \psi - \frac{1}{8\cdot 3!} \iota_\xi \check{c} \bar{\psi}\gamma^3 \vf\psi \\
         &\qquad \hspace{3mm}   -\frac{i}{4\cdot 3!} \bar{\psi}\gamma^3 \vf\left( \ugam\alpha(\check{c}\iota_\xi e \psi) \right)  + \frac{1}{2\cdot 3!}\bar{\psi}\gamma^3\chi\kappa\left[ <e, \bar{\chi}\left( -\frac{i}{2}\ugam^2\psi_0^\dag - [\check{\omega},\gamma]\psi - \frac{1}{2}\ugam\iota_\xi\check{c}\psi - \iota_\xi\ugam \check{c}\psi \right) > \right]\\
         &\qquad \hspace{3mm} + \frac{1}{16\cdot 3!}\bar{\psi}\gamma^3\chi \bar{\chi}\igam\igam\left(  -\frac{i}{2}\ugam^2\psi_0^\dag - [\check{\omega},\gamma]\psi - \frac{1}{2}\ugam\iota_\xi\check{c}\psi - \iota_\xi\ugam \check{c}\psi  \right)\\
        &\mathbb{q}_\psi = \frac{i}{4}\vf(\ugam \psi^0_\dag) -\frac{i}{4}\vf\left( \ugam\alpha( \check{\omega}\psi )  \right)   - \frac{i}{4}\vf\left( \ugam\alpha( \check{c}\iota_\xi e \psi )   \right) + \frac{1}{8}\vf\check{c} \chi - \frac{1}{8} \vf(\iota_\xi\check{c}\psi )\\
        & \hspace{7mm} +\frac{i}{4}\chi\kappa\left( <\bar{e}, \bar{\chi}\ugam^2\psi^0_\dag +i \bar{\chi}[\check{\omega}-\frac{i}{2}\iota_\xi \check{c} e + \iota_\xi e \check{c}]> \right)+\frac{1}{16}\chi\bar{\chi}\igam\igam(\ugam^2\psi^0_\dag +i [\check{\omega}-\frac{i}{2}\iota_\xi \check{c} e + \iota_\xi e \check{c}])\\
        &\frac{e^2}{2} \mathbb{q}_c = -\frac{i}{8}\bar{\chi}\vf\psi_\dag -\frac{i}{8\cdot 3!}\vf(\check{\omega }\bar{\chi}\gamma^3 \psi) -\frac{1}{2}\iota_\xi e \vf e^\dag+ \frac{1}{4}\iota_\xi( e \vf e^\dag)- \frac{i}{4\cdot 3!}\vf(\bar{\psi}^0_\dag \ugam) \gamma^3 \iota_\xi e \psi  \\
        &\qquad \hspace{3mm} + \frac{i}{4\cdot 3!}\vf\left(\alpha(\check{\omega}\bar{\psi}) \ugam\right) \gamma^3 \iota_\xi e \psi -\frac{i}{8}\iota_\xi(\bar{\psi} \vf\psi_\dag) - \frac{1}{8\cdot 3!}\iota_\xi(\check{\omega}\bar{\psi}\gamma^3 \vf\psi)\\
        &\qquad \hspace{3mm}+\frac{1}{4\cdot3!} \iota_\xi\left(\bar{\psi}\gamma^3\chi<e, \bar{\chi}([\check{\omega},\gamma]\psi + i\ugam^2\psi^0_\dag) >  \right)-\frac{1}{2\cdot 3!}\iota_\xi e \bar{\psi}\gamma^3\chi\kappa\left(<e, \bar{\chi}([\check{\omega},\gamma]\psi + i\ugam^2\psi^0_\dag) > \right)\\
        &\qquad \hspace{3mm} + \frac{1}{32\cdot 3!}\iota_\xi e \bar{\psi}\gamma^3\chi \bar{\chi}\igam\igam([\check{\omega},\gamma]\psi + i\ugam^2\psi^0_\dag) - \frac{1}{32\cdot3!} e \iota_\xi\left(\bar{\psi}\gamma^3\chi \bar{\chi}\igam\igam([\check{\omega},\gamma]\psi + i\ugam^2\psi^0_\dag)\right),
    \end{align*}
while the full vector field $Q$ is obtained by summing $Q=Q_0+\mathbb{q}$, $Q\chi=Q_0\chi$ and $Q\xi=Q_0\xi.$ 

Now, to keep the discussion somewhat contained, we explicitly compute $Q^2 e$ and show it vanishes, as similar computations and arguments work for the other fields and ghosts too.

Before we begin, we remark that $e\mathbb{q}_{\check{\omega}}=-\check{\omega}\mathbb{q}_e$ and $\frac{e^2}{2}\mathbb{q}_{\check{c}}=-e\mathbb{q}_e \check{c}$.
We then start by computing $\mathbb{q}(\varphi)=\bar{\chi}\gamma^a\chi \mathbb{q}(e_\mu^a)\partial_\mu$, obtaining
    \begin{align*}
        \mathbb{q}(\varphi^\mu)&=-e^\nu_a e^\mu_b \bar{\chi}\gamma^a\chi \mathbb{q}(e_\nu^b)\\
        &= -e^\mu_b \bar{\chi}\gamma^\nu \chi \left( \frac{1}{2}(\vf\check{\omega})^b_\nu  -\frac{1}{2}(\vf\check{c})_\nu \iota_\xi e^b -\frac{1}{4}(\vf(e^b\iota_\xi\check{c}))_\nu \right)\\
        &=-e^\mu_b \left( \frac{1}{2}\vf\vf\check{\omega}^b -\frac{1}{2}\vf\vf\check{c}\iota_\xi e^b -\frac{1}{4}\vf\vf(e^b\iota_\xi\check{c})  \right)=0,
    \end{align*}
since $\iota_\varphi$ is odd. Now we have
    \begin{align*}
        Q^2 e = Q_0^2 e + Q_0 \mathbb{q} e + \mathbb{q} Q_0 e + \mathbb{q}^2 e.
    \end{align*}
Notice that $\mathbb{q}^2 e$ is quadratic in the anti--fields, while the other terms are at most linear, hence we proceed to show  $\mathbb{q}^2 e=0$ separately. Notice first that from lemma \ref{lem: useful isos}.\ref{lem: iso W2 (1,1)}, we can equivalently compute $\frac{e^2}{2}Q^2 e$, obtaining
    \begin{align*}
        \frac{e^2}{2}\mathbb{q}^2 e=&\frac{e^2}{2}\mathbb{q} \left(  \frac{1}{2}\iota_\varphi\check{\omega} -\frac{1}{2}\iota_\varphi\check{c}\iota_\xi e -\frac{1}{4}\iota_\varphi(e \iota_\xi \check{c} )  \right)\\
        =&\frac{e^2}{2}\bigg[ -\frac{1}{2}\vf\mathbb{q}_{\check{\omega}} +\frac{1}{2}\vf\mathbb{q}_{\check{c}}\iota_\xi e +\frac{1}{2}\vf\check{c}\iota_\xi\left( \frac{1}{2}\iota_\varphi\check{\omega} -\frac{1}{2}\iota_\varphi\check{c}\iota_\xi e -\frac{1}{4}\iota_\varphi(e \iota_\xi \check{c} )\right)\\
        &\hspace{6.5mm} +\frac{1}{4}\vf(e \iota_\xi\mathbb{q}_{\check{c}} )  +\frac{1}{4}\vf\left(\iota_\xi\check{c}\left(  \frac{1}{2}\iota_\varphi\check{\omega} -\frac{1}{2}\iota_\varphi\check{c}\iota_\xi e -\frac{1}{4}\iota_\varphi(e \iota_\xi \check{c} )  \right)\right)\bigg]\\
        =&-\frac{e}{4}\vf(e\mathbb{q}_{\check{\omega}})+\frac{1}{4}\bar{\chi}\gamma\chi e\mathbb{q}_{\check{\omega}} - \frac{1}{4}\bar{\chi}\gamma\chi\iota_\xi\left(\frac{e^2}{2}\mathbb{q}_{\check{c}}\right)+\frac{e}{4}\vf\iota_\xi\left(\frac{e^2}{2}\mathbb{q}_{\check{c}}\right)\\
        &+\frac{e^2}{2}\bigg[ \frac{1}{4}\vf\check{c}\iota_\xi\vf\check{\omega} -\frac{1}{4} \vf\check{c}\iota_\xi\vf\check{c}\iota_\xi e +\frac{1}{8}\bar{\chi}\gamma\chi\vf\check{c}\iota_\xi\iota_\xi\check{c} -\frac{1}{8}\vf\check{c}\vf\iota_\xi\check{c} \iota_\xi e\\
        &\hspace{10mm}+\frac{1}{4}\iota_\xi\vf\check{c}\left(  \frac{1}{2}\iota_\varphi\check{\omega} -\frac{1}{2}\iota_\varphi\check{c}\iota_\xi e -\frac{1}{4}\iota_\varphi(e \iota_\xi \check{c} )  \right)\bigg].
    \end{align*}
Making the expressions containing $\mathbb{q}_{\check{c}}$ and $\mathbb{q}_{\check{\omega}}$ explicit is quite a cumbersome challenge. The reader will excuse us for not providing all the steps, however, when the dust settles, we are left with
    \begin{align*}
        \frac{e^2}{2}\mathbb{q}^2e=&\frac{1}{16}\bar{\chi}\gamma\chi\vf \left( \check{\omega}\iota_\xi(e\check{c}) -\check{\omega}^2 - (\iota_\xi e)^2 \check{c}^2+\frac{1}{4}\iota_\xi(e^2)\iota_\xi(\check{c}^2) + e^2(\iota_\xi \check{c})^2 \right)\\
        =&-\frac{1}{16}\vf\left[ \bar{\chi}\gamma\chi\left( \check{\omega}\iota_\xi(e\check{c}) -\check{\omega}^2 - (\iota_\xi e)^2 \check{c}^2+\frac{1}{4}\iota_\xi(e^2)\iota_\xi(\check{c}^2) + e^2(\iota_\xi \check{c})^2 \right)\right]=0.
    \end{align*}
To show that $\frac{e^2}{2}\mathbb{q}^2 e=0$, first consider any $\Xi\in \Omega^{(4,2})$, then the expression above is of the type $\bar{\chi}\gamma\chi \Xi\in\Omega^{(4,3)}$. Now, thanks to lemma \ref{lem: useful isos}.\ref{lem: iso rho (3,4)} and \ref{lem: iso (1,0) (3,4)}, we can see that there must exist a $\tilde{\theta}\in\Omega^{(1,0)}(\mathbb{S}_M)$ such that 
\begin{equation*}
    \bar{\chi}\gamma \Xi\chi= \frac{1}{3!}[e,e\bar{\chi}\gamma^3\ugam  \tilde{\theta}].
\end{equation*}
Similarly, using \ref{lem: useful isos}.\ref{lem: iso rho (0,1)}, there must exist a $\theta\in \Omega^{(0,1)} $ such that $\tilde{\theta}=[e,\theta]$, hence finding
    \begin{align*}
        \vf(\bar{\chi}\gamma\chi\Xi)&=\frac{1}{3!}\vf[e,e\bar{\chi}\gamma^3 \ugam \tilde{\theta}]\\
        &=\frac{1}{3!}[\bar{\chi}\gamma\chi, e \bar{\chi}\gamma^3\ugam\tilde{\theta}]-\frac{1}{3!}\underbrace{[e,\bar{\chi}\gamma^a\chi\bar{\chi}\gamma^3\gamma_a\tilde{\theta}]}_{0 \text{ from \eqref{id: Fierz identiy}}} + \frac{1}{3!}[e,e\bar{\chi}\gamma^3\ugam[\bar{\chi}\gamma\chi,\theta]]\\
        &=-\frac{1}{3!}[e,[\bar{\chi}\gamma\chi,e\bar{\chi}\gamma^3\ugam]]\theta=-\frac{1}{3!}[e,[\bar{\chi}\gamma\chi,e\bar{\chi}\gamma^3\ugam\theta^b]]v_b.
    \end{align*}
As it turns out, after some manipulation involving a mixture of Leibniz rule and Fierz identities, we have $[e,[\bar{\chi}\gamma\chi,e\bar{\chi}\gamma^3\ugam\theta^b]]=3[e,\bar{\chi}\gamma\chi\bar{\chi}\gamma^2\ugam^2\theta^b]$, while, thanks to \eqref{id: action of omega}, 
    \begin{align}\label{eq: Fierz di sto c..zo}
        \bar{\chi}\gamma\chi\bar{\chi}\gamma^2\ugam^2\theta^b&= \frac{1}{3}\bar{\chi}\gamma^3[\bar{\chi}\gamma^2\ugam^2\theta^b,\chi]=\frac{1}{3!}\bar{\chi}\gamma^3\gamma^a\gamma^c \chi \bar{\chi}\gamma_c\gamma_a \ugam^2\theta^b\\
        &\overset{\eqref{id: Fierz identiy}}{=}-\frac{1}{3!}\bar{\chi}\gamma^3\gamma^a\ugam^2\theta^b \bar{\chi}\gamma_a\chi\overset{\eqref{id: Fierz identiy}}{=}-\frac{1}{3}\bar{\chi}\gamma^3\gamma^a\chi \bar{\chi}\gamma^a\ugam^2\theta^b\nonumber\\
        &\overset{\eqref{flip:3}}{=}\frac{1}{3}\bar{\chi}\gamma^3\gamma^a\chi \bar{\theta}^b\ugam^2\gamma^a\chi\overset{\eqref{id: Fierz identiy}}{=}\frac{1}{3!}\bar{\chi}\gamma^3\gamma^a\ugam^2\theta^b \bar{\chi}\gamma_a\chi \nonumber\\
        &\overset{\eqref{eq: Fierz di sto c..zo}}{=} -\bar{\chi}\gamma\chi\bar{\chi}\gamma^2\ugam^2\theta^b=0,\nonumber
    \end{align}
hence showing $\frac{e^2}{2}\mathbb{q}^2 e=0$. 

We now consider the remaining terms of $Q^2 e$, which, after some rearranging, read
\begin{equation}\label{eq: Q^2 e}
\begin{split}
    &\hspace{-3mm}\mathbb{q}\left( \mathrm{L}_\xi^\omega e - [c,e]c+ \bar{\chi}\gamma\psi  \right) + Q_0\left( \frac{1}{2}\vf\check{\omega} -\frac{1}{2}\vf\check{c}\iota_\xi e -\frac{1}{4}\vf(\iota_\xi c e) \right)=\\
    =&[\iota_\xi \mathbb{q}_\omega,e] -[\mathbb{q}_c,e] -\bar{\chi}\gamma\mathbb{q}_\psi - \frac{1}{2}\vf(Q_0\check{\omega}) +\frac{1}{2}\vf(Q_0 \check{c}\iota_\xi e ) +\frac{1}{4}\vf(e\iota_\xi Q_0\check{c}) \\
    &+\frac{1}{2}\vf\mathrm{L}_\xi^\omega\check{\omega} -\frac{1}{2}\vf\mathrm{L}_\xi^\omega\check{c}\iota_\xi e -\frac{1}{4}\vf \check{c}\iota_{[\xi,\xi]}e -\frac{1}{8}\vf\iota_{[\xi,\xi]}\check{c} e + \frac{1}{8}\iota_{[\xi,\xi]}\check{c}\bar{\chi}\gamma\chi\\
    &-\frac{1}{4}e\vf\iota_\xi \mathrm{L}_\xi^\omega\check{c} +\frac{1}{4}\iota_\xi\mathrm{L}_\xi^\omega\check{c} \bar{\chi}\gamma\chi -\frac{1}{2}\vf[c,\check{\omega}] -\frac{1}{8}\bar{\chi}\gamma\chi\vf\check{c} +\frac{1}{2}\vf\check{c}\bar{\chi}\gamma\iota_\xi \psi \\
    &+\frac{1}{4}\vf\iota_\xi\check{c} \bar{\chi}\gamma\psi -\frac{1}{4}\iota_\xi\check{c}\bar{\chi}\gamma\vf\psi.
\end{split}
\end{equation}
A few remarks are in order. First of all, notice that the term $-\frac{1}{2}\vf Q_0\check{\omega}$ contains a term (proportional to the equations of motion) that cancels out exactly the non zero part of $Q_0^2 e$. Secondly, we notice that, in order to obtain $Q^2e=0$, we need to implement some terms in $\mathbb{q}_c$ to balance out $\iota_\xi\mathbb{q}_\omega$, in particular we are missing all the terms proportional to $\check{c}$. Explicitly, $e\mathbb{q}_\omega$ contains
    \begin{align}\label{eq: full elambda}
        \nonumber e\mathbb{l}(\check{c},\xi,\chi,\psi)\coloneqq&- \frac{1}{8\cdot 3!} \iota_\xi \check{c} \bar{\psi}\gamma^3 \vf\psi  -\frac{i}{4\cdot 3!} \bar{\psi}\gamma^3 \vf\left( \ugam\alpha(\check{c}\iota_\xi e \psi) \right)\\
        & -\frac{1}{2\cdot 3!}\bar{\psi}\gamma^3\chi\kappa\left[ <e, \bar{\chi}\left(  \frac{1}{2}\ugam\iota_\xi\check{c}\psi + \iota_\xi\ugam \check{c}\psi \right) > \right]\\
        &\nonumber- \frac{1}{16\cdot 3!}\bar{\psi}\gamma^3\chi \bar{\chi}\igam\igam\left(  \frac{1}{2}\ugam\iota_\xi\check{c}\psi + \iota_\xi\ugam \check{c}\psi  \right),
    \end{align}
hence, to cancel them in the computation of $Q_0^2e $ (and in general $Q_0^2$), we need to add\footnote{Similarly to adding $\iota_\xi \delta_\chi\omega$ to $Q_0c$} to $\mathbb{q}_c$ terms of the kind $\iota_\xi \mathbb{l}(\check{c},\xi,\varphi,\psi)$, resulting in a correction term in $s_2$
    \begin{equation*}
        \frac{1}{2}c^\dag\iota_\xi \mathbb{l}(\check{c},\xi,\varphi,\psi).
    \end{equation*}
Hence, in the computation of $\frac{e^2}{2}\big( [\iota_\xi\mathbb{q}_{\omega},e] -[\mathbb{q}_c,e] \big)$ we are left with
    \begin{align*}
        \frac{e^2}{2}\big( [\iota_\xi\mathbb{q}_{\omega},e] -[\mathbb{q}_c,e] \big)=&[e,\frac{e^2}{2}\mathbb{q}_c]+ \frac{1}{2}[e,e\mathbb{q}_\omega\iota_\xi e ]-\frac{1}{2}[e,e\iota_\xi(e\mathbb{q}_\omega)] \\
        =&\frac{e}{16\cdot 3!}[\iota_\xi\left(  \vf \check{c} \bar{\chi}\gamma^3\psi \right),e ]-\frac{1}{16\cdot 3!}[\iota_\xi e \vf\check{c}\bar{\chi}\gamma^3\psi,e]\\
        &-\frac{i}{8}[\bar{\chi}\vf\psi_\dag,e]+\frac{1}{8\cdot3!}[\vf(\check{\omega}\bar{\chi}\gamma^3\psi),e].
    \end{align*}
    
Similarly, when computing $-\frac{1}{2}\vf Q_0\check{\omega}+\frac{1}{2}\vf Q_0\check{c}\iota_\xi e + \frac{1}{4}\vf(e\iota_\xi Q_0\check{c})$, we notice that a lot of the terms in $Q_0\check{\omega}$ are canceled out by $\iota_\xi Q_0\check{c}$.\footnote{All the terms in $Q_0\check{\omega}$ coming from the variation in $\mathcal{S}_1$ of $\mathrm{L_\xi^\omega}(\cdot)$  with respect of $\omega$ are exactly canceled by the ones in $\iota_\xi Q_0\check{c}$ coming from the variation of $[c,\cdot]$ with respect to $c$} In particular, after noticing that 
    \begin{align*}
        -\frac{e^2}{4}\vf Q_0\check{\omega}=-\frac{e}{4}\vf(eQ_0\check{\omega}) + \frac{1}{4}\bar{\chi}\gamma\chi e Q_0\check{\omega}
    \end{align*}
and
    \begin{align*}
        \frac{e^2}{2}\left(\frac{1}{2}\vf Q_0\check{c}\iota_\xi e + \frac{1}{4}\vf(e\iota_\xi Q_0\check{c})\right)= \frac{e}{4}\vf\iota_\xi\left(\frac{e^2}{2}Q_0\check{c}\right)-\frac{1}{4}\bar{\chi}\gamma\chi\iota_\xi\left(\frac{e^2}{2}Q_0\check{c}\right),
    \end{align*}
one finds that the remaining terms are
\begin{equation}
    \begin{split}\label{eq: s}
        \frac{e^2}{2}&\left(-\frac{1}{2}\vf Q_0\check{\omega}+\frac{1}{2}\vf Q_0\check{c}\iota_\xi e + \frac{1}{4}\vf(e\iota_\xi Q_0\check{c})\right)=\\
        =&-\frac{e^2}{4}\vf\mathrm{EoM}_\omega +\unl{\frac{e}{4}\vf d_\omega\iota_\xi \omega^\dag}{s.1} -\unl{\frac{1}{4}\bar{\chi}\gamma\chi d_\omega\iota_\xi\omega^\dag}{s.2} - \frac{e^2}{4}[c,\vf\check{\omega}]-\unl{\frac{e}{8}\vf(d_\omega\iota_\xi\iota_\xi c^\dag + \iota_\xi\iota_\xi d_\omega c^\dag)}{s.3}\\
        &+\unl{\frac{1}{8}\bar{\chi}\gamma\chi(d_\omega\iota_\xi\iota_\xi c^\dag + \iota_\xi\iota_\xi d_\omega c^\dag)}{s.4}+\unl{\frac{e}{4}\vf(\check{\omega}\mathrm{L}_\xi^\omega e )}{s.5} -\unl{ \frac{1}{4}\bar{\chi}\gamma\chi \check{\omega}\mathrm{L}_\xi^\omega e}{s.6} + \frac{e}{4}\vf\left( \frac{1}{2}\bar{\chi}\gamma\psi\check{\omega} + \frac{1}{2\cdot3!} \bar{\chi}[\check{\omega},\gamma^3]\psi \right)\\
        &-\frac{1}{4}\bar{\chi}\gamma\chi \left( \frac{1}{2}\bar{\chi}\gamma\psi\check{\omega} + \frac{1}{2\cdot3!} \bar{\chi}[\check{\omega},\gamma^3]\psi \right)+\frac{e}{4}\vf\left[ \frac{1}{2}\bar{\chi}\gamma\psi \left(\iota_\xi e \check{c} +\frac{e}{2}\iota_\xi \check{c} \right) -\frac{1}{2\cdot 3!}\bar{\chi}[\left(\iota_\xi e \check{c} +\frac{e}{2}\iota_\xi \check{c} \right),\gamma^3]\psi \right]\\
        &-\frac{1}{4}\bar{\chi}\gamma\chi\left[ \frac{1}{2}\bar{\chi}\gamma\psi \left(\iota_\xi e \check{c} +\frac{e}{2}\iota_\xi \check{c} \right) -\frac{1}{2\cdot 3!}\bar{\chi}[\left(\iota_\xi e \check{c} +\frac{e}{2}\iota_\xi \check{c} \right),\gamma^3]\psi \right]\\
        &-\frac{e}{4}\vf\left( \frac{i}{2}e\bar{\psi}^0_\dag\ugam\gamma\chi -\frac{1}{2\cdot 3!}\bar{\psi}^0_\dag\ugam[e,\gamma^3]\chi \right) +\frac{1}{4}\bar{\chi}\gamma\chi\left( \frac{i}{2}e\bar{\psi}^0_\dag\ugam\gamma\chi -\frac{1}{2\cdot 3!}\bar{\psi}^0_\dag\ugam[e,\gamma^3]\chi \right)\\
        &-\unl{\frac{e}{4}\vf\iota_\xi\left(\frac{1}{2}\mathrm{L}_\xi^\omega(e^2)\check{c}\right)}{s.7} + \unl{\frac{1}{4}\bar{\chi}\gamma\chi \iota_\xi\left(\frac{1}{2}\mathrm{L}_\xi^\omega(e^2)\check{c}\right)}{s.8}-\frac{e}{4}\vf\iota_\xi(\bar\chi\gamma\psi e\check{c}) +\frac{1}{4}\bar{\chi}\gamma\chi\iota_\xi(\bar\chi\gamma\psi e\check{c})\\
        &-\unl{\frac{e}{4}\vf\iota_\xi d_\omega\omega^\dag }{s.9}+ \unl{\frac{1}{4}\iota_\xi d_\omega \omega^\dag}{s.10},
    \end{split}
\end{equation}
where we added terms proportional to $\iota_\xi\iota_\xi d_\omega c^\dag$, which vanish since $d_\omega c^\dag=0$. 

Now we notice that 
    \begin{itemize}
        \item $\reft{eq: s}{s.1}+\reft{eq: s}{s.2}+\reft{eq: s}{s.5}+\reft{eq: s}{s.6}+\reft{eq: s}{s.9}+\reft{eq: s}{s.10}=-\frac{e^2}{2}\vf\mathrm{L}_\xi^\omega \check{\omega}$;
        \item using the identity \cite{CCS2020}
            \begin{equation*}
                \frac{1}{2}\iota_{[\xi,\xi]} A= -\frac{1}{2}\iota_\xi\iota_\xi d_\omega A + \iota_\xi d_\omega \iota_\xi A - \frac{1}{2}d_\omega\iota_\xi \iota_\xi A 
            \end{equation*}
            and the fact that $\mathrm{L}_\xi\omega c^\dag= -d_\omega\iota_\xi c^\dag$, then
            \begin{align*}
                &\reft{eq: s}{s.3}+\reft{eq: s}{s.4}+\reft{eq: s}{s.7}+\reft{eq: s}{s.8}=\\
                &=\frac{e^2}{2}\left( \frac{1}{4}\vf \check{c} \iota_{[\xi,\xi]}e +\frac{1}{8}e\vf\iota_{[\xi,\xi]}\check{c} +\frac{1}{2}\iota_\xi e \vf\mathrm{L}_\xi^\omega\check{c} +\frac{1}{8}\bar{\chi}\gamma\chi \iota_{[\xi,\xi]}\check{c}+\frac{1}{4}\bar{\chi}\gamma\chi \iota_\xi \mathrm{L}_\xi^\omega\check{c} + \frac{e}{4}\vf\iota_\xi \mathrm{L}_\xi^\omega\check{c} \right).
            \end{align*}    
    \end{itemize}
Now we can finally compute the full $Q^2 e$, taking into consideration the full expression of $\bar{\chi}\gamma \mathbb{q}_\psi$, from \eqref{eq: Q^2 e}
    \begin{equation}
    \begin{split}
        Q^2 e =& \unl{\frac{e}{16\cdot 3!}[\iota_\xi\left(  \vf \check{c} \bar{\chi}\gamma^3\psi \right),e ]}{1} - \unl{\frac{1}{16\cdot 3!}[\iota_\xi e \vf\check{c}\bar{\chi}\gamma^3\psi,e]}{2} - \unl{\frac{i}{8}[\bar{\chi}\vf\psi_\dag,e]}{3} + \unl{\frac{1}{8\cdot3!}[\vf(\check{\omega}\bar{\chi}\gamma^3\psi),e]}{4}\\
        &- \unl{\frac{e^2}{2}\vf\mathrm{L}_\xi^\omega \check{\omega}}{5} - \unl{\frac{e^2}{4}[c,\vf\check{\omega}]}{6} + \unl{\frac{e}{4}\vf\left( \frac{1}{2}\bar{\chi}\gamma\psi\check{\omega} + \frac{1}{2\cdot3!} \bar{\chi}[\check{\omega},\gamma^3]\psi \right)}{7}\\
        & + \frac{e^2}{2}\left( \unl{\frac{1}{4}\vf \check{c} \iota_{[\xi,\xi]}e}{8} + \unl{\frac{1}{8}e\vf\iota_{[\xi,\xi]}\check{c}}{9} +\unl{\frac{1}{2}\iota_\xi e \vf\mathrm{L}_\xi^\omega\check{c}}{10} + \unl{\frac{1}{8}\bar{\chi}\gamma\chi \iota_{[\xi,\xi]}\check{c}}{11} + \unl{\frac{1}{4}\bar{\chi}\gamma\chi \iota_\xi \mathrm{L}_\xi^\omega\check{c}}{12} + \unl{\frac{e}{4}\vf\iota_\xi \mathrm{L}_\xi^\omega\check{c}}{13} \right)\\
        &-\unl{\frac{1}{4}\bar{\chi}\gamma\chi \left( \frac{1}{2}\bar{\chi}\gamma\psi\check{\omega} + \frac{1}{2\cdot3!} \bar{\chi}[\check{\omega},\gamma^3]\psi \right)}{14} + \frac{e}{4}\vf\left[ \unl{\frac{1}{2}\bar{\chi}\gamma\psi \left(\iota_\xi e \check{c} + \frac{e}{2}\iota_\xi \check{c} \right)}{15} - \unl{\frac{1}{2\cdot 3!}\bar{\chi}[\left(\iota_\xi e \check{c} +\frac{e}{2}\iota_\xi \check{c} \right)}{16},\gamma^3]\psi \right]\\
        &-\frac{1}{4}\bar{\chi}\gamma\chi\left[ \unl{\frac{1}{2}\bar{\chi}\gamma\psi \left(\iota_\xi e \check{c} +\frac{e}{2}\iota_\xi \check{c} \right)}{17} - \unl{\frac{1}{2\cdot 3!}\bar{\chi}[\left(\iota_\xi e \check{c} +\frac{e}{2}\iota_\xi \check{c} \right)}{18},\gamma^3]\psi \right]\\
        &-\unl{\frac{e}{4}\vf\left( \frac{i}{2}e\bar{\psi}^0_\dag\ugam\gamma\chi -\frac{1}{2\cdot 3!}\bar{\psi}^0_\dag\ugam[e,\gamma^3]\chi \right)}{19} +\unl{\frac{1}{4}\bar{\chi}\gamma\chi\left( \frac{i}{2}e\bar{\psi}^0_\dag\ugam\gamma\chi -\frac{1}{2\cdot 3!}\bar{\psi}^0_\dag\ugam[e,\gamma^3]\chi \right)}{20}\\
        &-\unl{\frac{e}{4}\vf\iota_\xi(\bar\chi\gamma\psi e\check{c})}{21} +\unl{\frac{1}{4}\bar{\chi}\gamma\chi\iota_\xi(\bar\chi\gamma\psi e\check{c})}{22} - \frac{e^2}{2}\bar{\chi}\gamma\left[\unl{ \frac{i}{4}\vf(\ugam \psi^0_\dag)}{23} -\unl{\frac{i}{4}\vf\left( \ugam\alpha( \check{\omega}\psi )  \right)}{24}   -\unl{ \frac{i}{4}\vf\left( \ugam\alpha( \check{c}\iota_\xi e \psi )   \right)}{25} \right]\\
        &-\frac{e^2}{2}\bar{\chi}\gamma\left[ \unl{\frac{1}{8}\vf\check{c} \chi}{26} - \unl{\frac{1}{8} \vf(\iota_\xi\check{c}\psi )}{27}+\frac{i}{4}\chi\kappa\left( <\bar{e}, \unl{\bar{\chi}\ugam^2\psi^0_\dag}{28} +i \bar{\chi}[\unl{\check{\omega}}{29}-\unl{\frac{i}{2}\iota_\xi \check{c} e }{30}+ \unl{\iota_\xi e \check{c}}{31},\gamma]\psi> \right)\right]\\
        &-\frac{e^2}{2}\bar{\chi}\gamma\left(\frac{1}{16}\chi\bar{\chi}\igam\igam(\unl{\ugam^2\psi^0_\dag}{32} +i [\unl{\check{\omega}}{33}-\unl{\frac{i}{2}\iota_\xi \check{c} e}{34} + \unl{\iota_\xi e \check{c}}{35},\gamma]\psi)\right)+\frac{e^2}{2}\left(\unl{ \frac{1}{4}\vf\iota_\xi\check{c} \bar{\chi}\gamma\psi}{36} -\unl{\frac{1}{4}\iota_\xi\check{c}\bar{\chi}\gamma\vf\psi }{37}\right)\\
        &+\frac{e^2}{2}\left(\unl{\frac{1}{2}\vf\mathrm{L}_\xi^\omega\check{\omega}}{38} -\unl{\frac{1}{2}\vf\mathrm{L}_\xi^\omega\check{c}\iota_\xi e }{39}-\unl{\frac{1}{4}\vf \check{c}\iota_{[\xi,\xi]}e }{40}-\unl{\frac{1}{8}\vf\iota_{[\xi,\xi]}\check{c} e }{41}+\unl{ \frac{1}{8}\iota_{[\xi,\xi]}\check{c}\bar{\chi}\gamma\chi}{42}\right)\\
        &+\frac{e^2}{2}\left(-\unl{\frac{1}{4}e\vf\iota_\xi \mathrm{L}_\xi^\omega\check{c}}{43} +\unl{\frac{1}{4}\iota_\xi\mathrm{L}_\xi^\omega\check{c} \bar{\chi}\gamma\chi}{44} -\unl{\frac{1}{2}\vf[c,\check{\omega}] }{45}-\unl{\frac{1}{8}\bar{\chi}\gamma\chi\vf\check{c}}{46} +\unl{\frac{1}{2}\vf\check{c}\bar{\chi}\gamma\iota_\xi \psi }{47}\right).
    \end{split}
    \end{equation}    
We can regroup the above terms to show that the total sum is zero. We immediately see
\begin{itemize}
    \item $\reft{eq: Q^2 e}{5} + \reft{eq: Q^2 e}{38}=0$,
    \item $ \reft{eq: Q^2 e}{6} + \reft{eq: Q^2 e}{45}=0 $,
    \item $ \reft{eq: Q^2 e}{8} + \reft{eq: Q^2 e}{40}=0 $,
    \item $\reft{eq: Q^2 e}{9} + \reft{eq: Q^2 e}{41} = 0$,
    \item $\reft{eq: Q^2 e}{10} + \reft{eq: Q^2 e}{39} = 0$,
    \item $\reft{eq: Q^2 e}{11} + \reft{eq: Q^2 e}{42} = 0$,
    \item $\reft{eq: Q^2 e}{12} + \reft{eq: Q^2 e}{44} = 0$,
    \item $\reft{eq: Q^2 e}{13} + \reft{eq: Q^2 e}{43} = 0$,
    \item $\reft{eq: Q^2 e}{26} + \reft{eq: Q^2 e}{46} = 0$,
\end{itemize}
For the remaining terms there is a recurring pattern which we explicitly show just once. Consider for example $\reft{eq: Q^2 e}{3} + \reft{eq: Q^2 e}{19} +\reft{eq: Q^2 e}{20} + \reft{eq: Q^2 e}{23} + \reft{eq: Q^2 e}{28} + \reft{eq: Q^2 e}{32} $, we have, after expanding the terms
\begin{itemize}
    \item $\reft{eq: Q^2 e}{3}=\frac{i}{8}e \left( \frac{1}{2}\bar{\chi}\gamma^2\ugam\vf(\ugam\psi^0_\dag ) + e \bar{\chi}\gamma\vf(\ugam\psi^0_\dag) \right)$,
    \item $\reft{eq: Q^2 e}{19} + \reft{eq: Q^2 e}{20} = \frac{i}{16}e\vf( \bar{\psi}^0_\dag \ugam )\ugam\gamma^2\chi + \frac{i}{8\cdot 3!}e \bar{\psi}^0_\dag\ugam[\bar{\chi}\gamma\chi,\gamma^3]\chi - \frac{i}{8\cdot 3!}\bar{\chi}\gamma\chi \bar{\psi}^0_\dag\ugam[e,\gamma^3]\chi $,
    \item the term $\reft{eq: Q^2 e}{28} + \reft{eq: Q^2 e}{32}$ presents an added difficulty, which can be resolved once one notices that, backwards engineering the methods used to compute $\delta_\chi^2 \psi$ in the previous section,\footnote{In this particular case, it suffices to notice that $\reft{eq: Q^2 e}{28} + \reft{eq: Q^2 e}{32}$ is exactly equal to 
    \begin{equation*}
        -\frac{e^2}{4}\bar{\chi}\gamma[\alpha,\chi],
    \end{equation*}
    which is easily seen after comparing it with the expression of $[\delta_\chi\omega,\chi]$. One then just substitutes $d_\omega\psi$ in $\delta_\chi\omega$ with $i\ugam^2\psi^0_\dag$ to find that $\alpha$ is such that $e\alpha=\frac{i}{3!}\bar{\chi}\gamma^3\ugam\psi^0_\dag$. Then the above expression becomes 
        \begin{align*}
            -\frac{e^2}{4}\bar{\chi}\gamma[\alpha,\chi]=-\frac{e^2}{8}[\alpha,\bar{\chi}\gamma\chi].
        \end{align*}
    } it can be rewritten as
        \begin{align*}
            \reft{eq: Q^2 e}{28} + \reft{eq: Q^2 e}{32}&=\frac{e^2}{8}\left[(W_1^{(2,3)})^{-1}\left( \frac{i}{3!}\bar{\chi}\gamma^3\ugam\psi^0_\dag \right),\bar{\chi}\gamma\chi\right]\\
            &= -\frac{e}{8}\left[\bar{\chi}\gamma\chi, \frac{i}{3!}\bar{\chi}\gamma^3\ugam\psi^0_\dag \right] -\frac{i}{8\cdot 3!}\bar{\chi}\gamma^3\ugam\psi^0_\dag[e,\bar{\chi}\gamma\chi]\\
            &\overset{\ref{lem: Fierz}}{=}-\frac{e}{8}\left[\bar{\chi}\gamma\chi, \frac{i}{3!}\bar{\chi}\gamma^3\ugam\psi^0_\dag \right] + \frac{i}{8\cdot 3!}\bar{\chi}[e,\gamma^3]\ugam\psi^0_\dag\bar{\chi}\gamma\chi.
            \end{align*}
    it is a simple matter of algebra to see $\reft{eq: Q^2 e}{3} + \reft{eq: Q^2 e}{19} +\reft{eq: Q^2 e}{20} + \reft{eq: Q^2 e}{23} + \reft{eq: Q^2 e}{28} + \reft{eq: Q^2 e}{32}=0$. 
\end{itemize}
As previously anticipated, one can analogously show the following terms vanish
\begin{itemize}
    \item $\reft{eq: Q^2 e}{1} +\reft{eq: Q^2 e}{2}+ \reft{eq: Q^2 e}{2} + \reft{eq: Q^2 e}{15} + \reft{eq: Q^2 e}{16} +\reft{eq: Q^2 e}{17} +\reft{eq: Q^2 e}{18} +\reft{eq: Q^2 e}{21} + \reft{eq: Q^2 e}{22} + \reft{eq: Q^2 e}{25} +\reft{eq: Q^2 e}{27} + \reft{eq: Q^2 e}{30} + \reft{eq: Q^2 e}{31} + \reft{eq: Q^2 e}{34} +\reft{eq: Q^2 e}{35} + \reft{eq: Q^2 e}{36} +\reft{eq: Q^2 e}{37} +\reft{eq: Q^2 e}{47}         =0$
    \item $ \reft{eq: Q^2 e}{4} + \reft{eq: Q^2 e}{7} +\reft{eq: Q^2 e}{14} + \reft{eq: Q^2 e}{29} + \reft{eq: Q^2 e}{33}=0 $
\end{itemize}

Now, in order to show that $Q^2=0$ when computed on the other fields and ghosts, one needs to perform similar manipulations as in the case of $Q^2 e$, but we think that explicitly carrying them out, while equally (if not more) challenging, does not provide any further insight.

\nocite{*}
\newrefcontext[sorting=nty]
\sloppy
\printbibliography

\end{document}